\documentclass{article}

\usepackage{amsfonts,amssymb}
\usepackage{amsmath}
\usepackage{epsfig}
\usepackage{float}
\usepackage{indentfirst}
\usepackage{color}
\usepackage{caption}
\usepackage{graphicx}

\usepackage{multirow}
\usepackage[table]{xcolor}
\usepackage{rotating}

\numberwithin{equation}{section}
\begin{document}

\title{Active cloaking of finite defects for flexural waves in elastic plates}

\author{J. O'Neill$^1$,  \"O. Selsil$^1$, R.C. McPhedran$^{1,2}$, A.B. Movchan$^1$ and N.V. Movchan$^1$, \\
$^1$ Department of Mathematical Sciences, University of Liverpool,
 Liverpool L69 7ZL, UK,\\
 $^2$ CUDOS ARC Centre of Excellence, School of Physics, University of Sydney, Sydney, \\New South Wales 2006, Australia.}

\maketitle


\begin{abstract}
We present a new method to create an active cloak for a rigid inclusion in a thin plate, and analyse flexural waves within such a plate governed by the Kirchhoff plate equation.
We consider scattering of both a plane wave and a cylindrical wave  by a single clamped inclusion of circular shape. 
In order to cloak the inclusion, we place control sources at small distances from the scatterer and choose their intensities to eliminate propagating orders of the scattered wave, thus reconstructing the respective incident wave.
We then vary the number and position of the control sources to obtain the most effective configuration for cloaking the circular inclusion. Finally, we successfully cloak an arbitrarily shaped scatterer in a thin plate by deriving a semi-analytical, asymptotic algorithm.
\end{abstract}

\section{Introduction}
Wave propagation in doubly-periodic perforated thin plates has been recently extensively studied, stimulated by the work of Movchan {\it et al.}  \cite{ABM_NVM_RCM},
where they assumed circular geometry for inclusions, which were either clamped or free. They constructed an analytical solution of the spectral problem for the biharmonic operator using the method of multipole expansions. Interestingly, it was discovered that for the clamped boundary case, there existed a ``platonic" band gap for the zero radius limit. In 2009, McPhedran {\it et al.} \cite{RCM_ABM_NVM} considered the same problem and explored the properties of clamped holes in the zero radius limit in more detail. It has important physical consequences  that the Green's function for the biharmonic operator, as discussed  by Evans \& Porter \cite{DVE_RP}, is non-singular at the source point.

More recently, Movchan {\it et al.} \cite{NVM_RCM_ABM_CGP} investigated flexural wave scattering by sets of periodic gratings, where they established a rigorous model to describe the interaction of incident plane waves and gratings of circular inclusions (or fixed point scatterers in the zero radius limit). Here, the incident plane wave was written in terms of Bessel function expansions for both Helmholtz and modified Helmholtz type waves and the solution for the flexural wave function was constructed in terms of both types of plane waves and multipole expansions.  

Coating objects  with appropriately designed or chosen materials for cloaking, starting with Wolf \& Habashy  \cite{EW_TH} and Nicorovici {\it et al.}  \cite{NAN_RCM_GWM}, is a well-established technique. Further studies by  Al\`u \& Engheta \cite{AA_NE1}, \cite{AA_NE2}, Leonhardt  \cite{UL}, Pendry {\it et al.} \cite{JBP_DS_DRS} and Milton {\it et al.} \cite{GWM_NAPN} are also notable amongst many others. Since the beginning of 2007 there has been a plethora of publications on cloaking; more than 200 articles mention cloaking in their titles, with articles in the thousands having been stimulated by the ground-breaking papers just cited. 

Lately, there has been a revival in the investigation of active cloaking, which relies on locating a number of active sources outside the object. For early work on this method we refer to Miller (2006) \cite{DABM} (active interior cloaking) and Nicorovici {\it et al.} (2007) \cite{NAPN_GWM_RCM_LCB}, which were followed by Guevara Vasquez {\it et al.} \cite{GVF_GWM_DO_2009a} and Zheng {\it et al.} \cite{HHZ_JJX_YL_CTC}, the latter using the method to create deceptive cloaking. Norris {\it et al.} (2012) \cite{ANN_FAA_WJP1} developed a method to evaluate the amplitudes of active source terms that made direct use of the results of Guevara Vasquez {\it et al.} \cite{GVF_GWM_DO_2011a}, which was later extended  to elastodynamic cloaking (Norris {\it et al.} (2014) \cite{ANN_FAA_WJP2}). 
Transformation elastodynamics, as well as active exterior acoustic cloaking, are the topics presented in great detail in the comprehensive article by Guevara Vasquez {\it et al.} \cite{GVF_GWM_DO_PS}. Our work here was stimulated
by the work of the Milton group (see for example Guevara Vasquez {\it et al.} \cite{GVF_GWM_DO_2009b}, \cite{GVF_GWM_DO_2011b}). One advantage of the active cloaking method over the metamaterial approach is that it is, in principle, easier  for the former to overcome
the inevitable bandwidth difficulties associated with the latter, which were recently emphasised by Chen {\it et al.} \cite{PYC_CA_AA}.

The flexural wave problem involves both Helmhotlz and modified Helmholtz waves in its solution. An important feature is that all modified Helmholtz waves are evanescent, and so their amplitudes
do not have to be controlled to deliver  good cloaking at large distances from the target object. These extra benign degrees of freedom in low order multipole terms are, we think, important to the excellent quality of the
cloaking we demonstrate in this article.

To the best of our knowledge, using the method of multipole expansions to successfully cloak an inclusion in a thin plate has not been previously studied. In this paper, we provide an analytical model first for the active cloaking of a circular shaped inclusion and then extend this to 
 an arbitrarily shaped inclusion for flexural waves in an infinite plate using the method of multipole expansion. We make use of the solution for bending waves in plates, introduced by Movchan \textit{et al.} \cite{ABM_NVM_RCM}, for a single inclusion rather than a square array of circular holes. We consider the scattering of both plane and cylindrical waves from the inclusion and discuss the improvement of the cloaking in detail.  

Section \ref{prob_form} deals with the formulation of the problem for the out-of-plane elastic displacement. The requirement to satisfy the clamped boundary conditions leads us to a set of equations connecting the various constants appearing in the solution to each other, via the scattering matrix.
In section \ref{Scatter_PlaneWave}, we give the general representation of the scattered field from an incident plane wave. We introduce two active control sources in section \ref{intro_control_sources}, to annul the coefficients of the monopole and dipole terms of the propagating part of the scattered field, thus defining the necessary equations for the evaluation of the source amplitudes. Numerical examples associated with two control sources are given in section \ref{Num_Egs_Two_Sources}, illustrating the effectiveness of the cloaking. In section \ref{cloaking_additional _control_sources}, we extend the method to include more control sources around the inclusion and present numerical results clearly indicating a notable improvement in the cloaking of the inclusion, with six sources producing effectively perfect cloaking. We also remark on the robustness of six-source cloaking to rotations. The analysis for an incident plane wave is reconsidered for a cylindrical wave from a point source in section \ref{Scatter_CircularWave}. In section \ref{arbitrarily_shaped}, we analyse 
active cloaking for flexural waves from a scatterer of arbitrary shape, and illustrate its effectiveness in a numerical example.

\section{Problem formulation}
\label{prob_form}

Our ultimate aim is to study wave propagation in a thin elastic plate with an arbitrarily shaped inclusion. For the sake of convenience, we first assume that the inclusion is of circular shape with radius $r_c$. The out-of-plane elastic displacement $W(\bf x)$ satisfies the equation of motion
\begin{equation}
\Delta^2 W ({\bf x}; t) + \frac{\rho h}{D} \ddot{W}({\bf x}; t) = 0, \quad {\bf x}=(x_1,x_2) \in \Omega \setminus D_{r_c}\label{gov_eqn}
\end{equation}
where $\Delta^2$ is the biharmonic operator, dot on the variable denotes the derivative with respect to time $t$, $\rho$ is the mass density, $h$ is the thickness of the plate,
$D_{r_c} = \{{\bf x}: x_1^2+x_2^2<r_c^2\}$ is the circle of radius $r_c$, $D=Eh^3/[12(1-\nu^2)]$ is the plate's flexural rigidity, with $E$ the Young's modulus and $\nu$ the Poisson's ratio of the elastic material.

Assuming time-harmonic vibrations, i.e. $W({\bf x}; t) = w(r,\theta) \,{\rm exp (i\omega t)}$, the governing equation (\ref{gov_eqn}) can be reduced to 
\begin{equation}
\Delta^2 w (r,\theta) - \frac{\rho h \omega^2}{D} w(r,\theta) = (\Delta + \beta^2)(\Delta - \beta^2)w (r,\theta)= 0, \label{gov_eqn_simp}
\end{equation}
here $\beta^2 = \omega\sqrt{\rho h / D}$ is the spectral parameter.

We write the solution of equation (\ref{gov_eqn_simp}) as
$$
w(r,\theta) = \sum_{n=-\infty}^\infty w_n(r,\theta),
$$
where
\begin{equation}
w_n(r,\theta)=[A_n J_n (\beta r)+E_n H_n^{(1)}(\beta r)+B_n I_n (\beta r)+F_n K_n (\beta r)]\, e^{i n\theta},
\label{gen_soln_nth_coeff}
\end{equation}
is the $n$th multipole component of the flexural wave disturbance.
We impose the conditions $w_n(r_c,\theta)=0$, $\partial w_n(r_c,\theta)/\partial r=0$ on the surface of the rigid cylinder. These conditions lead to the following
\begin{equation}
\left[ \begin{tabular}{c c}
$H_n^{(1)}(\beta r_c)$ & $K_n (\beta r_c)$\\
$H_n^{(1)'} (\beta r_c)$ & $K_n' (\beta r_c)$
\end{tabular}\right] \left[\begin{tabular}{c}
$E_n$\\
$F_n$
\end{tabular}\right]=-\left[ \begin{tabular}{c c}
$J_n (\beta r_c)$ & $I_n (\beta r_c)$\\
$J_n ' (\beta r_c)$ & $I_n' (\beta r_c)$
\end{tabular}\right] \left[\begin{tabular}{c}
$A_n$\\
$B_n$
\end{tabular}\right].
\label{gen_soln_satisfying_bcs}
\end{equation}

The solution of equation (\ref{gen_soln_satisfying_bcs}) may be written in terms of Wronskians, which we will denote by ${\cal W}[\cdot]$:
\begin{equation}
\begin{split}
{\cal W} [H_n^{(1)}(\beta r_c),K_n(\beta r_c)]&=H_n^{(1)}(\beta r_c) K_n' (\beta r_c)-H_n^{(1)'} (\beta r_c) K_n (\beta r_c),\\
{\cal W} [ K_n (\beta r_c),J_n(\beta r_c)]&=K_n (\beta r_c) J_n' (\beta r_c)-K_n '(\beta r_c) J_n (\beta r_c), \\
{\cal W} [K_n (\beta r_c),I_n(\beta r_c)]&=K_n (\beta r_c) I_n' (\beta r_c)-K_n '(\beta r_c) I_n (\beta r_c)=\frac{1}{\beta r_c},\\
{\cal W} [J_n (\beta r_c),H_n^{(1)}(\beta r_c)]&=J_n (\beta r_c) H_n^{(1)'} (\beta r_c)-J_n '(\beta r_c) H_n^{(1)} (\beta r_c)=\frac{2 i}{\pi\beta r_c}, \\
{\cal W} [I_n (\beta r_c),H_n^{(1)}(\beta r_c)]&=I_n (\beta r_c) H_n^{(1)'} (\beta r_c)-I_n '(\beta r_c) H_n^{(1)} (\beta r_c).
\label{Wronskians}
\end{split}
\end{equation}
The solution is then
\begin{equation}
\left[\begin{tabular}{c}
$E_n$\\
$F_n$
\end{tabular} \right]=\frac{1}{{\cal W} [H_n^{(1)}(\beta r_c),K_n(\beta r_c)]}\left[\begin{tabular}{c c}
${\cal W}[K_n (\beta r_c),J_n(\beta r_c)]$, & $1/(\beta r_c)$\\
$ 2 i/(\pi \beta r_c)$, & ${\cal W}[I_n (\beta r_c),H_n^{(1)}(\beta r_c)]$
\end{tabular}\right]  \left[\begin{tabular}{c}
$A_n$\\
$B_n$
\end{tabular}\right] .
\label{gen_solns_in_terms_of_Wronskians}
\end{equation}
We denote the entries of the $n$th order scattering matrix in (\ref{gen_solns_in_terms_of_Wronskians}) by ${\cal S}_n$. Small frequency expansions ($\beta\rightarrow 0$) of the scattering matrix elements for $n=0,\pm 1$ are given in the appendix.

Note that
for any $n \neq 0$, the scattering  matrix elements obey the following symmetry relations:
\begin{equation}
\begin{array}{cc}
{\cal S}_{-n}[1,1](\beta r_c)={\cal S}_{n}[1,1](\beta r_c), & {\cal S}_{-n}[1,2](\beta r_c)=(-1)^n{\cal S}_{n}[1,2](\beta r_c),\\ \\
{\cal S}_{-n}[2,1](\beta r_c)=(-1)^n{\cal S}_{n}[2,1](\beta r_c), &{\cal S}_{-n}[2,2](\beta r_c)={\cal S}_{n}[2,2](\beta r_c).
\label{S_symmetry_relations}
\end{array}
\end{equation}

\section{Scattering of a plane wave by a rigid inclusion}
\label{Scatter_PlaneWave}

We consider flexural plane waves travelling at an angle $\theta_0$ to the $x$-axis and interacting with a rigid inclusion placed with its centre at the origin. The incident field is
\begin{equation} 
w_0(x,y)=\exp \{i\beta (x\cos\theta_0+y\sin\theta_0)\},
\label{incident_plane_wave}
\end{equation}
or, using the Jacobi-Anger expansion, in polar form
\begin{equation}
w_0(r,\theta)=\exp \{i \beta r \cos (\theta-\theta_0)\}=\sum_{n=-\infty}^\infty i^n J_n(\beta r) e^{i n(\theta-\theta_0)}.
\label{incident_plane_wave_polar_form}
\end{equation}
The scattered flexural wave has its monopole and dipole terms given by
\begin{eqnarray}
w_{sc}(r,\theta)&=&{\cal S}_0[1,1](\beta r_c)H_0^{(1)}(\beta r)+{\cal S}_0[2,1](\beta r_c) K_0(\beta r) \nonumber \\
 && +2 i \cos (\theta-\theta_0)\left[{\cal S}_1[1,1](\beta r_c)H_1^{(1)}(\beta r)+{\cal S}_1[2,1](\beta r_c) K_1(\beta r)\right].
\label{scattered_flexural_wave_monopole_dipole}
\end{eqnarray}
Neglecting terms containing the factor $\beta^2$, we obtain
\begin{equation}
w_{sc}(r,\theta)=-\left[ H_0^{(1)}(\beta r)+\frac{2 i}{\pi}K_0(\beta r)  \right]-\frac{2 i \pi\cos (\theta-\theta_0)}{4 i\gamma+\pi+4 i\log(\beta r_c/2)}\left[ H_1^{(1)}(\beta r)+\frac{2 i}{\pi}K_1(\beta r)  \right].
\label{wsc_monopole_dipole_S_expanded}
\end{equation}

Using the symmetry relations (\ref{S_symmetry_relations}), the general expression for the scattered field is
\begin{eqnarray}
w_{sc}(r,\theta)&=&{\cal S}_0[1,1](\beta r_c)H_0^{(1)}(\beta r)+{\cal S}_0[2,1](\beta r_c) K_0(\beta r) \nonumber\\
 & &+\sum_{n=1}^\infty [{\cal S}_n[1,1](\beta r_c)H_n^{(1)}(\beta r)+{\cal S}_n[2,1](\beta r_c) K_n(\beta r)]i^n e^{i n (\theta-\theta_0)} \nonumber\\
  & &+\sum_{n=1}^\infty [{\cal S}_n[1,1](\beta r_c)(-1)^n H_n^{(1)}(\beta r)+(-1)^n{\cal S}_n[2,1](\beta r_c) K_n(\beta r)] (-i)^n e^{-i n (\theta-\theta_0)}. \nonumber \\
  & & 
\label{wsc_monopole_dipole_general}
\end{eqnarray}
Hence,
\begin{eqnarray}
w_{sc}(r,\theta)&=&{\cal S}_0[1,1](\beta r_c)H_0^{(1)}(\beta r)+{\cal S}_0[2,1](\beta r_c) K_0(\beta r) \nonumber\\
 & &+\sum_{n=1}^\infty 2 i^n [{\cal S}_n[1,1](\beta r_c)H_n^{(1)}(\beta r)+{\cal S}_n[2,1](\beta r_c) K_n(\beta r)] \cos  n (\theta-\theta_0) . \nonumber\\
\label{wsc_monopole_dipole_general_simplified}
\end{eqnarray}

\section{Scattering of a plane wave by a rigid inclusion with control sources present}
\label{intro_control_sources}

We next add two point sources of flexural waves at the points $(-a,0)$ and $(b,0)$. The amplitudes of these point sources are respectively $Q_-$ and $Q_+$; these are to be chosen to annul selected multipole coefficients in
the propagating part of the field scattered by the cylinder. 

The field incident on the cylinder now consists of the plane wave terms from the previous section, plus the multipole expansions of  the control sources. These are:
\begin{equation}
Q_+ G(x-b,y)=-\frac{Q_+}{8 \beta^2}\left[i H_0^{(1)}(\beta\sqrt{(x-b)^2+y^2})-\frac{2}{\pi}K_0(\beta\sqrt{(x-b)^2+y^2})\right],
\label{Qp_Greens_fn}
\end{equation}
or, using Graf's addition theorem,
\begin{equation}
Q_+ G(x-b,y)=-\frac{Q_+}{8 \beta^2}\sum_{l=-\infty}^\infty  e^{i l\theta}\left[i H_l^{(1)}(\beta b)J_l(\beta r)-\frac{2}{\pi} K_l(\beta b)I_l(\beta r)\right],
\label{Qp_Greens_fn_Grafs_addition}
\end{equation}
and
\begin{equation}
Q_- G(x+a,y)=-\frac{Q_-}{8 \beta^2}\left[i H_0^{(1)}(\beta\sqrt{(x+a)^2+y^2})-\frac{2}{\pi}K_0(\beta\sqrt{(x+a)^2+y^2})\right],
\label{Qm_Greens_fn}
\end{equation}
and so
\begin{equation}
Q_- G(x+a,y)=-\frac{Q_-}{8 \beta^2}\sum_{l=-\infty}^\infty  e^{i l(\pi-\theta)}\left[i H_l^{(1)}(\beta a)J_l(\beta r)-\frac{2}{\pi} K_l(\beta a)I_l(\beta r)\right].
\label{Qm_Greens_fn_Grafs_addition}
\end{equation}
The $n$th order coefficients for the total wave incident on the cylinder are then
\begin{equation}
\begin{split}
A_n&=i^n e^{-i n\theta_0}-i\frac{Q_+}{8\beta^2}H_n^{(1)}(\beta b)-i\frac{Q_-}{8\beta^2}(-1)^n H_n^{(1)}(\beta a),\\
B_n&= \frac{Q_+}{4\pi \beta^2}K_n(\beta b)+\frac{Q_-}{4\pi \beta^2}(-1)^n K_n(\beta a).
\label{An_Bn}
\end{split}
\end{equation}
From equation (\ref{gen_solns_in_terms_of_Wronskians}), we can write the following for the $n$th order coefficients for the outgoing waves from the cylinder as
\begin{equation}
E_n={\cal S}_n[1,1] A_n + {\cal S}_n[1,2] B_n,~~ F_n={\cal S}_n[2,1] A_n+{\cal S}_n[2,2] B_n.
\label{En_Fn}
\end{equation}
Thus, the $n$th order coefficient for the $H_n^{(1)} (\beta r)$  term in the field for $r>\max(a,b)$ is
\begin{equation}
{\cal S}_n[1,1] A_n + {\cal S}_n[1,2] B_n-\frac{i}{8\beta^2}[Q_+J_n(\beta b)+(-1)^n Q_- J_n(\beta a)],
\label{Hn_coeff}
\end{equation}
while that for $K_n(\beta r)$ is
\begin{equation}
{\cal S}_n[2,1] A_n + {\cal S}_n[2,2] B_n+\frac{1}{4\pi \beta^2}[Q_+ I_n(\beta b)+(-1)^n Q_- I_n(\beta a)].
\label{Kn_coeff}
\end{equation}

We choose the coefficients $Q_+$, $Q_-$ and the positions $b$, $a$ so that the monopole and dipole coefficients of the Hankel terms are zero. The equations to be satisfied are
\begin{equation}
{\cal S}_0[1,1]  A_0 + {\cal S}_0[1,2] B_0-\frac{i}{8\beta^2}[Q_+J_0(\beta b)+Q_- J_0(\beta a)]=0,
\label{H0_coeff_eq_0}
\end{equation}
\begin{equation}
{\cal S}_1[1,1] A_1 + {\cal S}_1[1,2]  B_1-\frac{i}{8\beta^2}[Q_+J_1(\beta b)-Q_- J_1(\beta a)]=0,
\label{H1_coeff_eq_0}
\end{equation}
and
\begin{equation}
{\cal S}_1[1,1] A_{-1 }-{\cal S}_1[1,2]  B_{-1}+\frac{i}{8\beta^2}[Q_+J_1(\beta b)-Q_- J_1(\beta a)]=0.
\label{Hneg1_coeff_eq_0}
\end{equation}
Using equation (\ref{An_Bn}), we find  for the monopole and dipole terms
\begin{equation}
S_0[1,1] = {\cal L}_0, \quad i e^{-i \theta_0} S_1[1,1]  ={\cal L}_1, \quad i e^{ i \theta_0} S_1[1,1]  ={\cal L}_1, 
\label{eqns_for_Q+Q-}
\end{equation}
where 
\begin{eqnarray}
{\cal L}_k &=& 
\frac{i Q_+}{8\beta^2}\bigg[{\cal S}_k[1,1]  H_k^{(1)}(\beta b)+\frac{2 i}{\pi} {\cal S}_k[1,2] K_k(\beta b)+J_k(\beta b)\bigg]\nonumber\\
 & &+(-1)^k\,\frac{i Q_-}{8\beta^2}\bigg[{\cal S}_k[1,1] H_k^{(1)}(\beta a)+\frac{2 i}{\pi} {\cal S}_k[1,2]  K_k(\beta a)+J_k(\beta a)\bigg], 
 \quad k=0,1. \label{L_k}
\end{eqnarray}

We note that the last two equations in (\ref{eqns_for_Q+Q-}) 
are equivalent if $\theta_0=0$ and that ${\cal L}_k  = {\cal L}_{-k} $ for any integer, $k$.

\section{Reduction of the shadow region by two control sources}
\label{Num_Egs_Two_Sources}

We now consider examples illustrating the effect of control sources on the scattering by a cylinder with the boundary conditions of zero wave amplitude and normal derivative on its surface.
The active cloaking is expected to be most effective in the case where $\beta$ is small: the dimensionless parameter $\beta r_c$ should be distinctly smaller than unity. For given $\beta$, $r_c$, $a$ and $b$,  and with $\theta_0=0$, the linear equations (\ref{eqns_for_Q+Q-})  are solved for $Q_-$ and $Q_+$. The $n$th  order outgoing wave coefficients $E_n$ and $F_n$ follow then from equations (\ref{An_Bn}) and (\ref{En_Fn}). Everywhere outside the scatterer, the flexural wave is given by
\begin{eqnarray}
w(r,\theta)&=&w_0(r,\theta)+Q_- G(r\cos \theta+a, r\sin \theta)+Q_+ G(r\cos \theta-b, r\sin \theta)\nonumber\\
& &+\sum_{n=-N}^N [E_n H_n^{(1)}(\beta r)+F_n K_n(\beta r)]\,e^{i n\theta},
\label{Wtotal}
\end{eqnarray}
where the summation limit $N$ is chosen to be sufficiently large to ensure accuracy of the wave amplitude. 

The expression (\ref{Wtotal}) of course gives a complex value, its real part representing the wave amplitude
at a particular instant of time. Animations can be made of $\Re[ \exp(-i\Phi) w(r,\theta)]$, with the phase $\Phi$ evolving from $0$ to $2\pi$, or figures constructed for a particular value of $\Phi$.

\begin{figure}[H]
\begin{center}
\includegraphics[width=6.5cm]{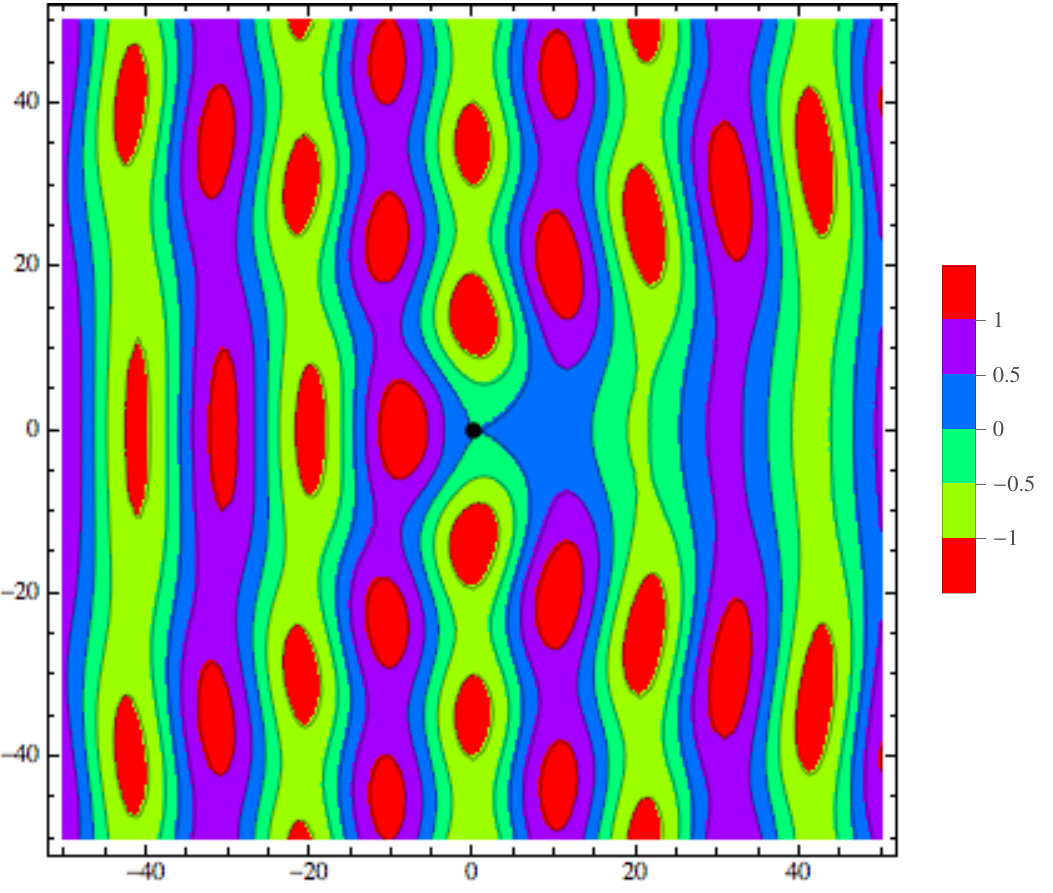} 
\includegraphics[width=6.5cm]{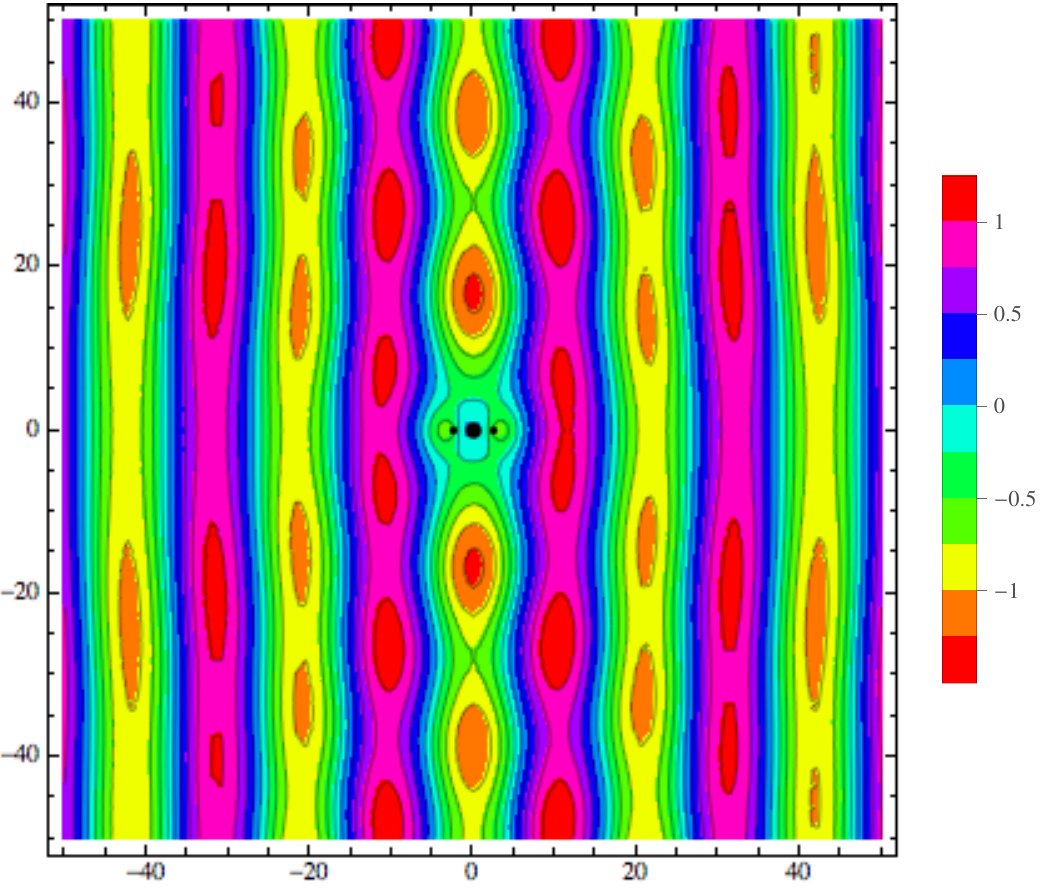}
\end{center}
\caption{In both figures $\beta=0.3$, $r_c=1.0$, $N=2$, $\Phi=\pi$. On the left, we show the flexural wave amplitude for a cylinder with no control sources where the black dot depicts the position of the cylinder. The right figure shows the flexural wave amplitude for a cylinder with two control sources. The larger black dot gives the position of the cylinder; the two smaller
black dots give the position of the two control sources: $a=b=2.5$, $Q_-=-3.075+1.728\,i$, $Q_+=-3.075-1.728\,i$.} \label{L_wave_amp_no_sources_R_wave_amp_2_sources}
\end{figure}

Figure \ref{L_wave_amp_no_sources_R_wave_amp_2_sources} (left) shows the flexural wave amplitude for a cylinder with no control sources present. The region of small amplitude behind the cylinder is the most prominent sign in the wave plot of the presence of the cylinder. Figure \ref{L_wave_amp_no_sources_R_wave_amp_2_sources} (right) shows the flexural amplitude pattern when two control sources with amplitudes $Q_-=-3.075+1.728\,i$ and $Q_+=-3.075-1.728\,i$ at the positions $(-a,0)$ and $(b,0)$, respectively, (with $a=b=2.5$), are used to annul the monopole and dipole outgoing wave amplitudes. One can see that the boundaries between regions of positive and negative amplitude are far straighter, and the indication of the scattering shadow has been suppressed; a maximum of vibration amplitude sits just behind the cylinder. Note that the required strengths of the control sources increase strongly as $a$ and $b$ tend towards $r_c$. (For example, with $a=b=1.5$, they are $Q_\mp=-19.4\pm 10.5\,i$.) The fact that the control amplitudes $Q_-$, $Q_+$ are complex indicates that the phase of these sources is an important parameter. 

\vspace{-0.15cm}

\begin{figure}[H]
\begin{center}
\includegraphics[width=6cm]{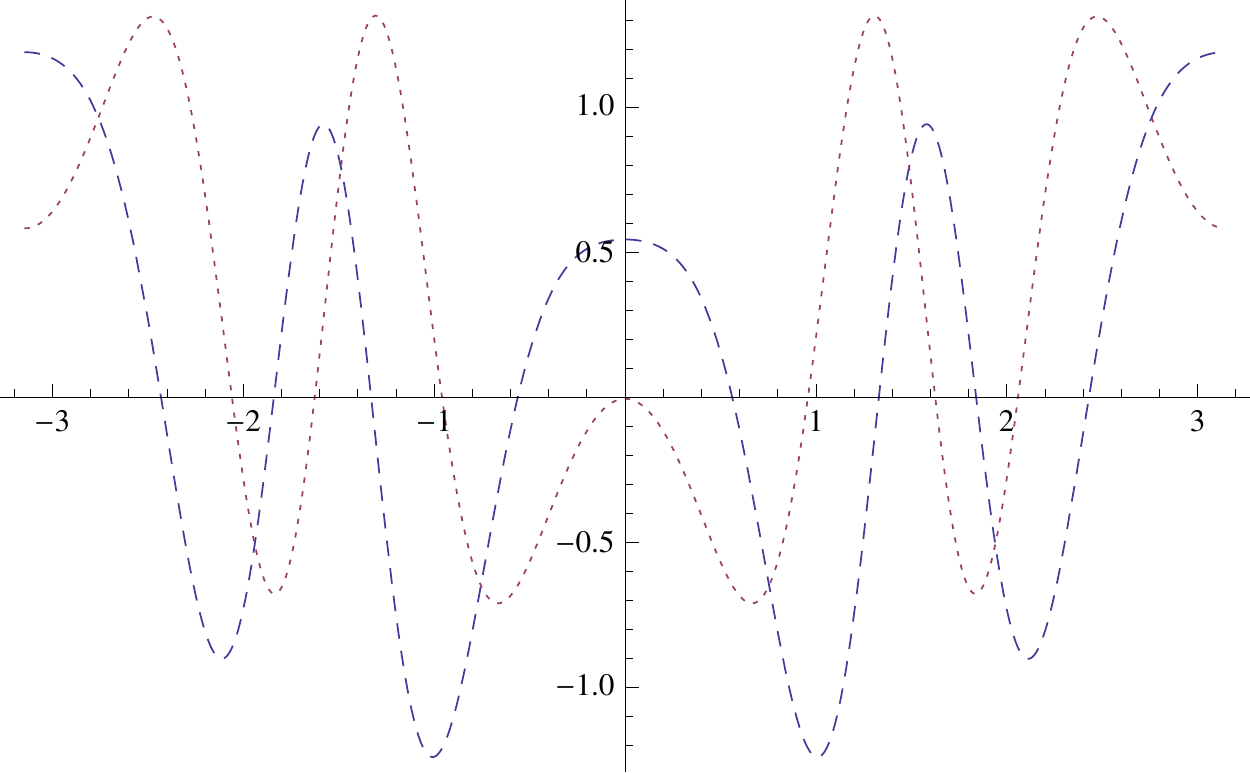}~~\includegraphics[width=6cm]{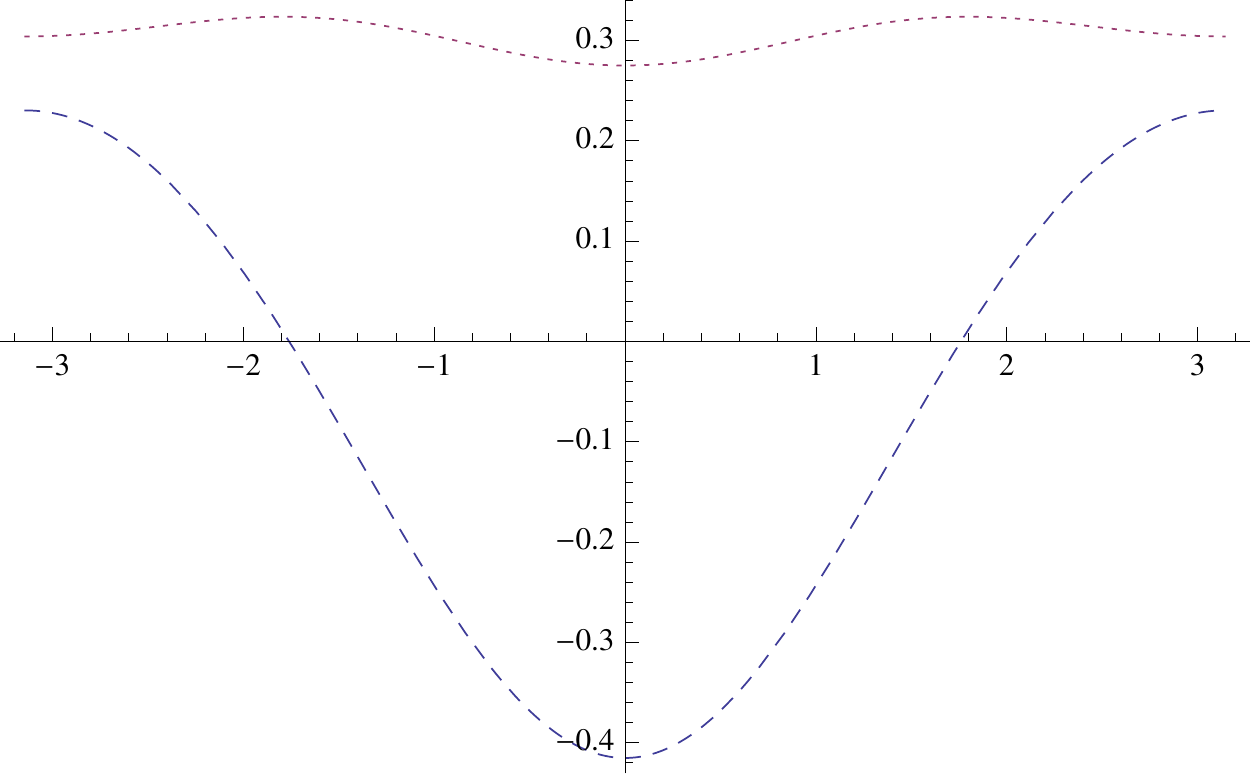}\\
 \includegraphics[width=6cm]{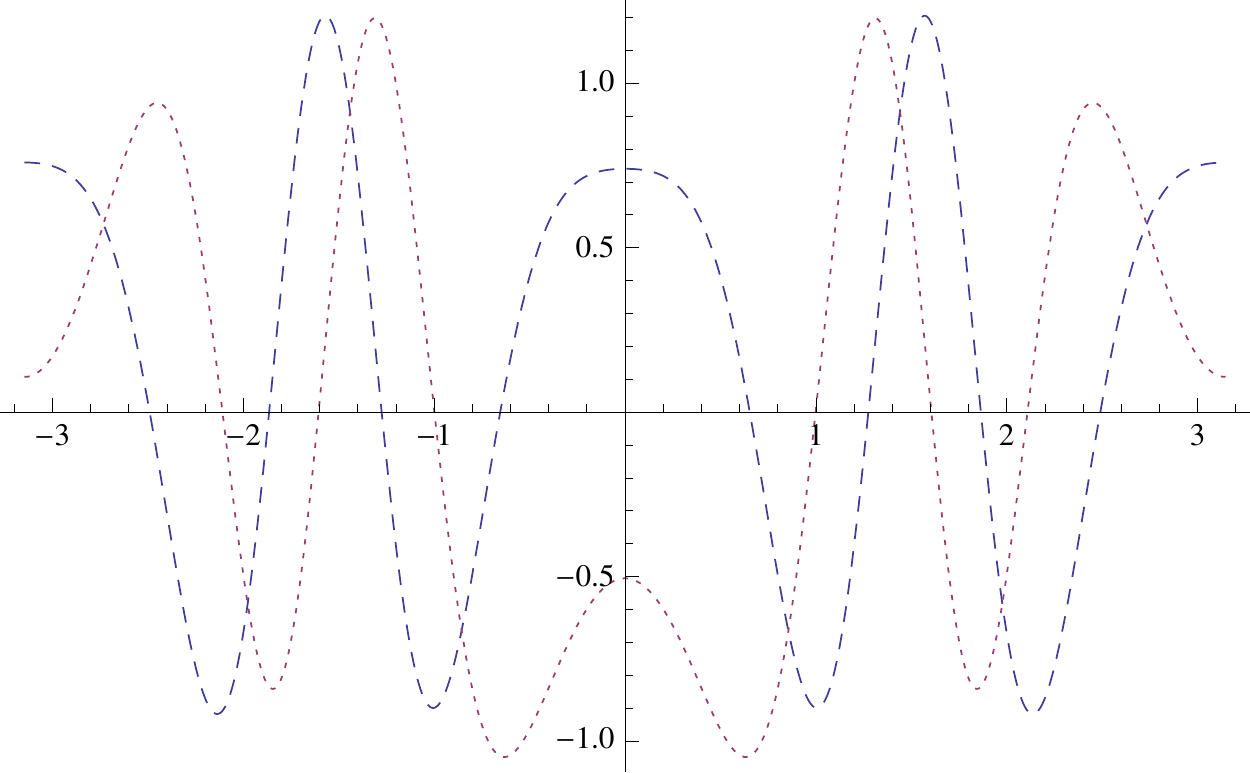}~~\includegraphics[width=6cm]{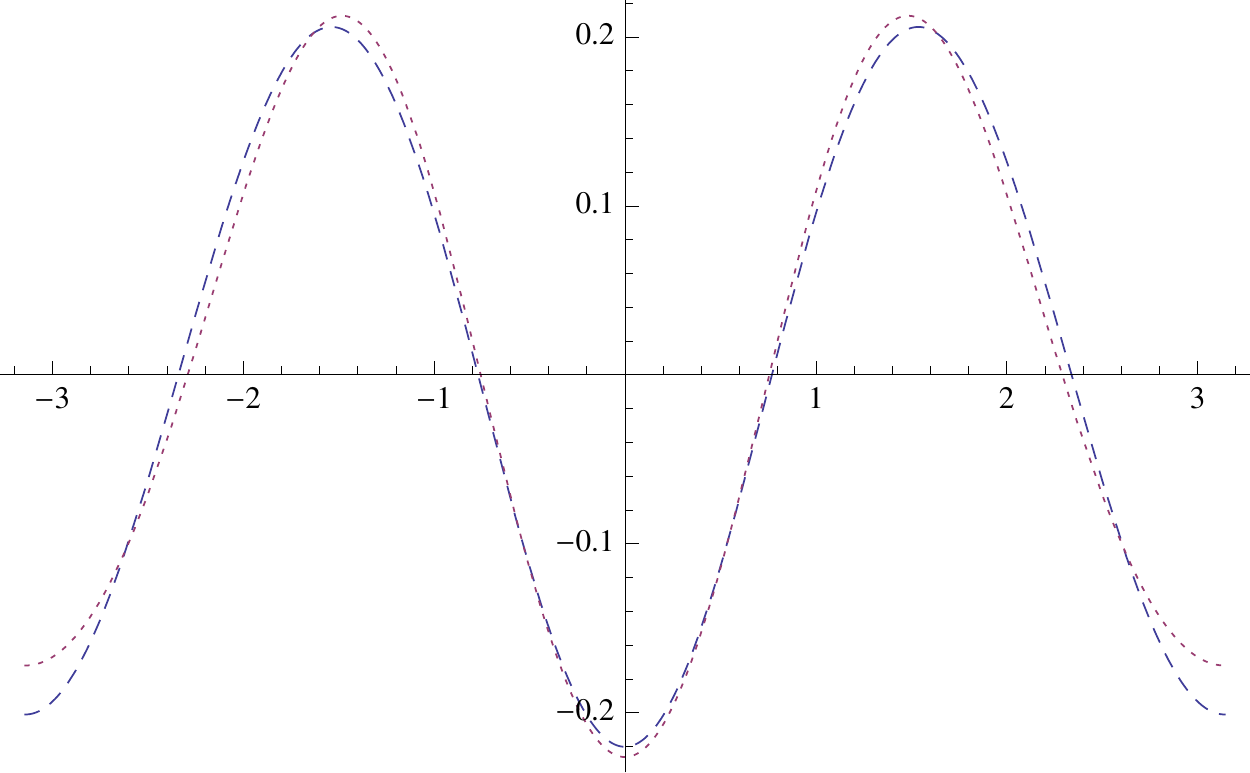}
  \end{center}
  \caption{The flexural wave amplitude on a circle of radius 20, as a function of $\theta$,  for a cylinder with no control sources (top) and with two control sources (bottom). On the left, we show the total amplitude, and on the right the scattered amplitude. In both figures $\beta=0.3$, $r_c=1.0$, $N=2$, blue (dashed): real part, red (dotted): imaginary part. For two control sources $a=b=2.5$, $Q_-=-3.075+1.728\,i$, $Q_+=-3.075-1.728\,i$. } 
\label{circular_plot_no_sources_and_two_sources}
\end{figure}

\vspace{-0.5cm}
Another way to show the flexural wave amplitude is to plot it around a circle, whose radius is sufficiently large for the modified Helmholtz terms to have died away. In figure \ref{circular_plot_no_sources_and_two_sources}, we compare the
angular variations with no control sources to those with two control sources. It is immediately clear from the right-hand plots that the control sources have successfully eliminated the monopole and dipole terms from the flexural wave, leaving the $\cos 2\theta$ like term as the leading contributor to the wave amplitude.

\begin{figure}[H]
\begin{center}
\includegraphics[width=6.5cm]{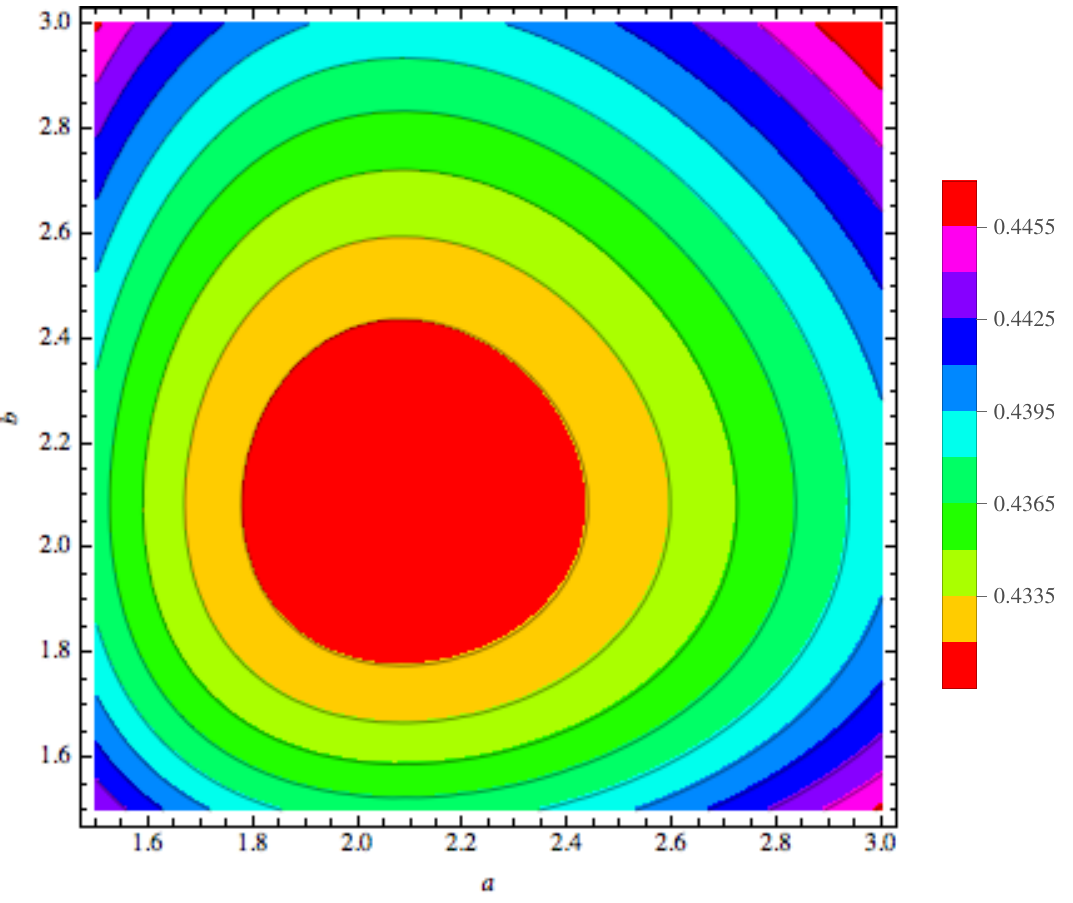}
\end{center}
\caption{The amplitude of the  outgoing Hankel function term with angular dependence $\exp( 2 i \theta)$  in the flexural wave  expansion, plotted as a function of the positions $a$ and
$b$ of the control sources. Note the almost constant amplitude levels (see the legend).} \label{Hn_term_angular_dependence}
\end{figure}

A natural query relates to the optimal choice of the positions of the control sources. Given that for any $a$ and $b$ one can make zero the monopole and dipole terms in the outgoing flexural wave, one might hope to minimise the amplitude of the quadrupole term by varying $a$ and $b$. In figure \ref{Hn_term_angular_dependence} we show that in fact these variables have little effect on the quadrupole amplitude, so that two control sources can only be used to control two angular components of the scattered flexural wave.

\vspace{-0.2cm}

\section{Active cloaking with additional control sources}
\label{cloaking_additional _control_sources}

The cloaking can be improved by increasing the number of control sources of flexural waves. We will assume $\theta_0=0$ throughout this section, and any further active sources are introduced off the $x$-axis, surrounding the inclusion. 
However, it is important to consider the implications on symmetry before adding more active sources.  For a plane wave incident along the $x$-axis, the scattering problem is symmetric under a change in $y$ for an inclusion whose profile is itself, symmetric.  In this case, the scattering problem is even in $y$, and any active control sources should be placed in order to preserve this symmetry. As a result, the amplitudes of certain control sources are also restricted; those located opposite each other, either side of the $x$-axis, are equal, inevitably reducing the number of unknown intensities to be calculated.  If symmetry is broken by the positioning of additional active sources, or by a scatterer which does not have a symmetric profile, then the number of unknown intensities to be calculated is equal to the number of active sources, with this being the case for an arbitrarily shaped scatterer.

The multipole expansions for two additional control sources at $(u\cos \xi,\pm u \sin \xi), \, 0< \xi<\pi/2$, are
\begin{eqnarray}
P \,G(x-u\cos \xi,y \mp u\sin \xi) &=& -\frac{P}{8\beta^2} \left[ i H_0^{(1)} (\beta\sqrt{(x-u\cos\xi)^2+(y \mp u \sin \xi)^2}) \right.\nonumber \\
& & \left.-\frac{2}{\pi}K_0 (\beta\sqrt{(x-u\cos\xi)^2+(y \mp u \sin \xi)^2})\right],
\end{eqnarray}
where $P$ denotes the amplitude of the pair. Graf's addition theorem then gives
\begin{equation}
P \,G(x-u\cos \xi,y \mp u\sin \xi) = 
-\frac{P}{8 \beta^2}\sum_{l=-\infty}^\infty  
e^{i l(\theta\mp\xi)}\left[i H_l^{(1)}(\beta u)J_l(\beta r)-\frac{2}{\pi} K_l(\beta u)I_l(\beta r)\right], \label{P_Greens_fn}
\end{equation}
leading to the $n$th order coefficients for the total wave incident on the cylinder as
\begin{equation}
\begin{split}
A_n &= i^n - i\frac{Q_+}{8\beta^2}H_n^{(1)}(\beta b)-i\frac{Q_-}{8\beta^2}(-1)^n H_n^{(1)}(\beta a)
- \frac{i P}{4\beta^2} H_n^{(1)} (\beta u)\, \cos (n\xi), \\
B_n &= \frac{Q_+}{4\pi \beta^2}K_n(\beta b)+\frac{Q_-}{4\pi \beta^2}(-1)^n K_n(\beta a) 
+ \frac{P}{2\pi\beta^2} K_n(\beta u) \, \cos (n\xi). \label{An_Bn_4sources}
\end{split}
\end{equation}
As in section \ref{Scatter_PlaneWave}, we set the $n$th order coefficient for the $H_n^{(1)}$ term equal to zero, but this time where $r>\max{(a,b,u)}$, to obtain the following constraints (compare with equation 
(\ref{eqns_for_Q+Q-}))
\begin{equation}
{\cal S}_0[1,1] = {\cal M}_0, \quad i  {\cal S}_1[1,1]  ={\cal M}_1, \quad 
- {\cal S}_2[1,1] = {\cal M}_2, \label{eqns_for_Q+Q-P}
\end{equation}
where 
\begin{equation}
{\cal M}_k = {\cal L}_k  +  \frac{i P}{4\beta^2} \left[{\cal S}_k[1,1]  H_k^{(1)}(\beta u)+\frac{2 i}{\pi} {\cal S}_k[1,2]  K_k(\beta u)
+J_k(\beta u)\right]\, \cos (k\xi), \quad k=0,1,2. \label{M_k}
\end{equation}
Note that  ${\cal L}_2$ is defined by the formula (\ref{L_k}) for $k=2$.

Once again, for given $\beta, r_c, a, b,u$ and $\xi$, we solve equations  (\ref{eqns_for_Q+Q-P}) for the unknown wave amplitudes $Q_\pm$ and $P$, then modify the representation for the flexural wave outside the scatterer as
\begin{eqnarray}
w(r,\theta)&=&w_0(r,\theta)+Q_- G(r\cos \theta+a, r\sin \theta)+Q_+ G(r\cos \theta-b, r\sin \theta)\nonumber\\[0.2cm] 
\vspace{0.2cm}
& &+P\, G(r\cos \theta-u \cos \xi, r\sin \theta-u \sin \xi)
+P\, G(r\cos \theta-u \cos \xi, r\sin \theta+u \sin \xi) \nonumber\\
& & +\sum_{n=-N}^N [E_n H_n^{(1)}(\beta r)+F_n K_n(\beta r)]\,e^{i n\theta}.
\label{Wtotal_add2}
\end{eqnarray}
Of course, the remark on the choice of $N$ from section \ref{Num_Egs_Two_Sources} is still valid (see equation (\ref{Wtotal})).

\begin{figure}[H]
\begin{center}
\includegraphics[width=6.5cm]{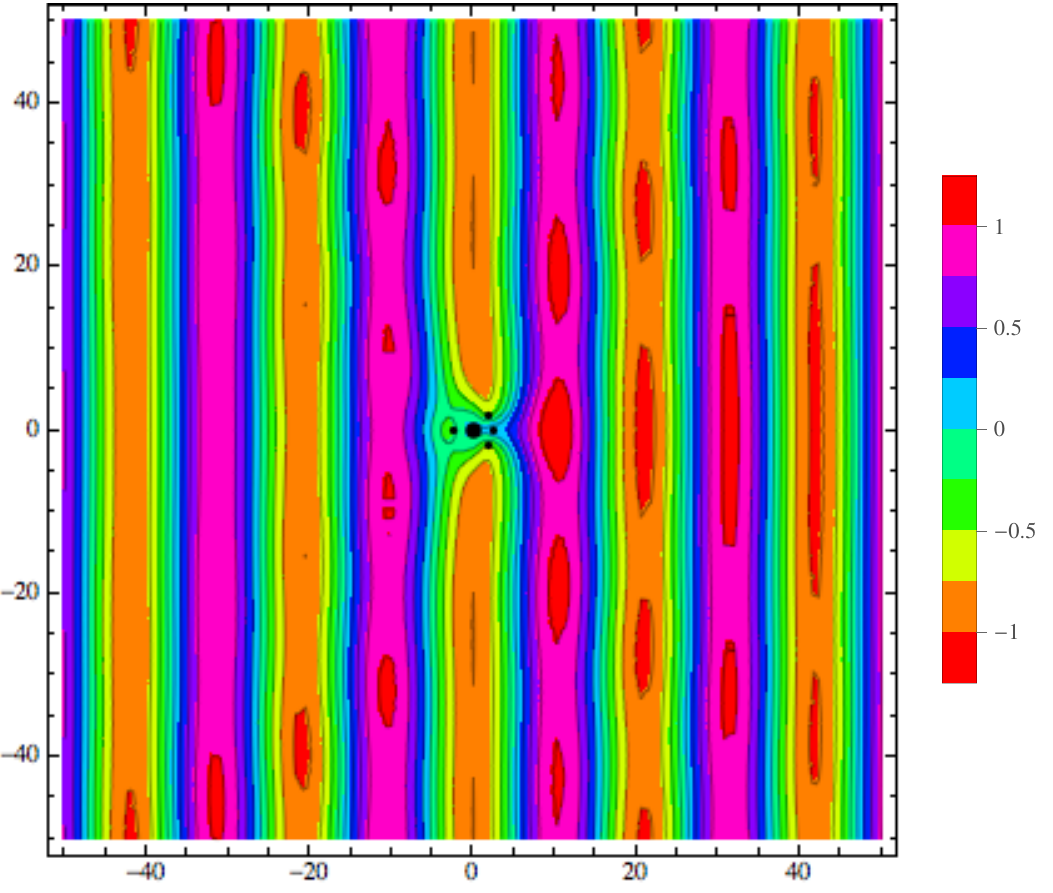}
\includegraphics[width=6.5cm]{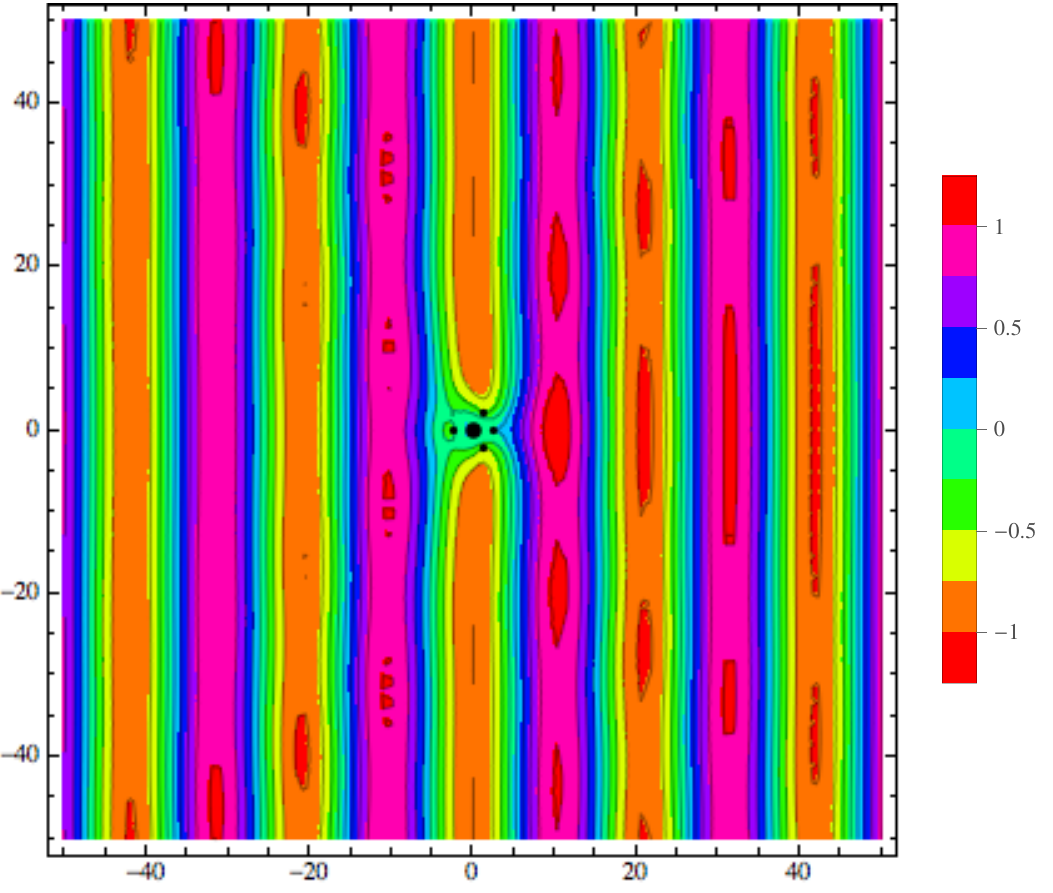} \\
\includegraphics[width=6.5cm]{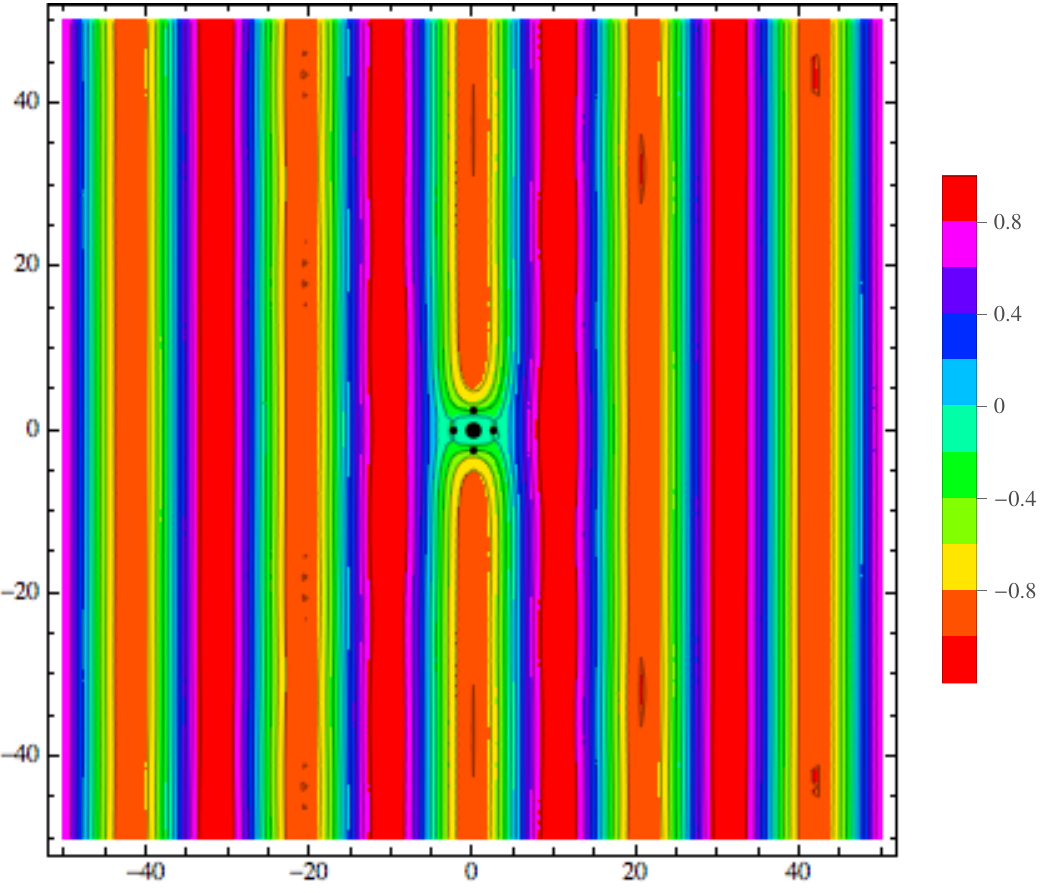} \\
\includegraphics[width=6cm]{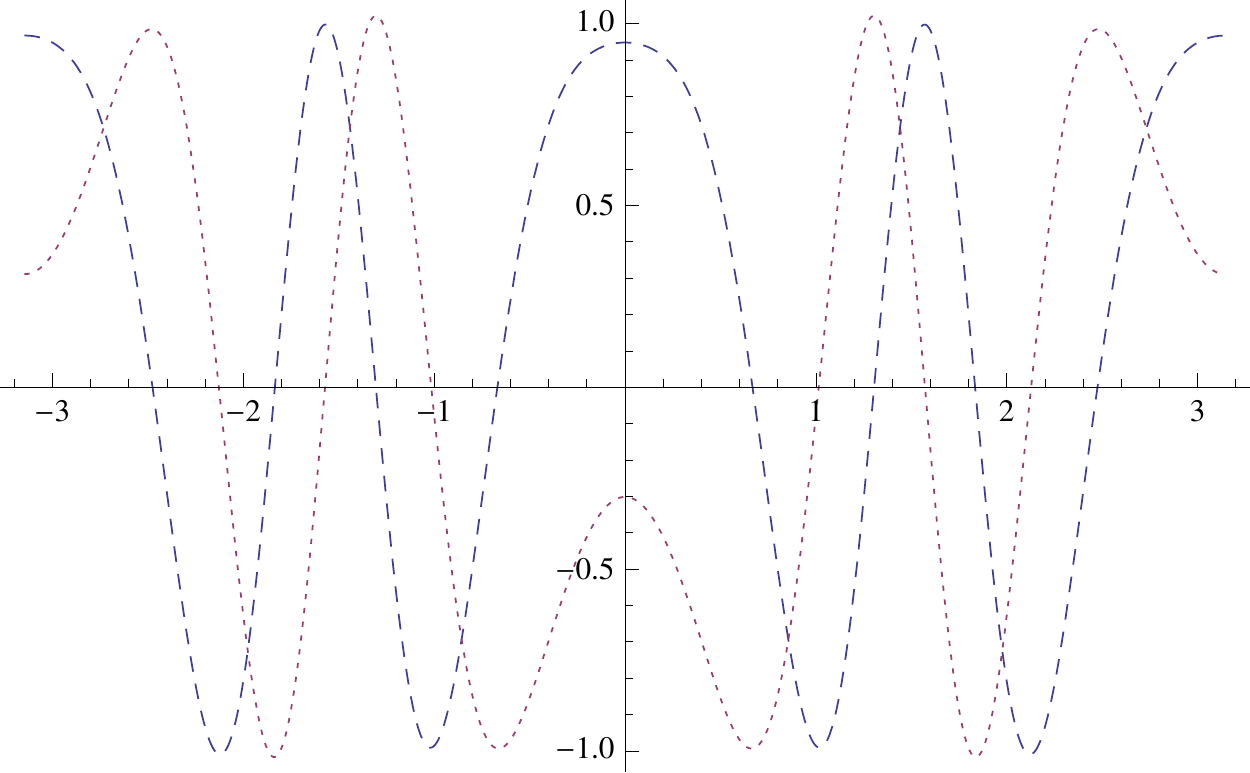}~~
\includegraphics[width=6cm]{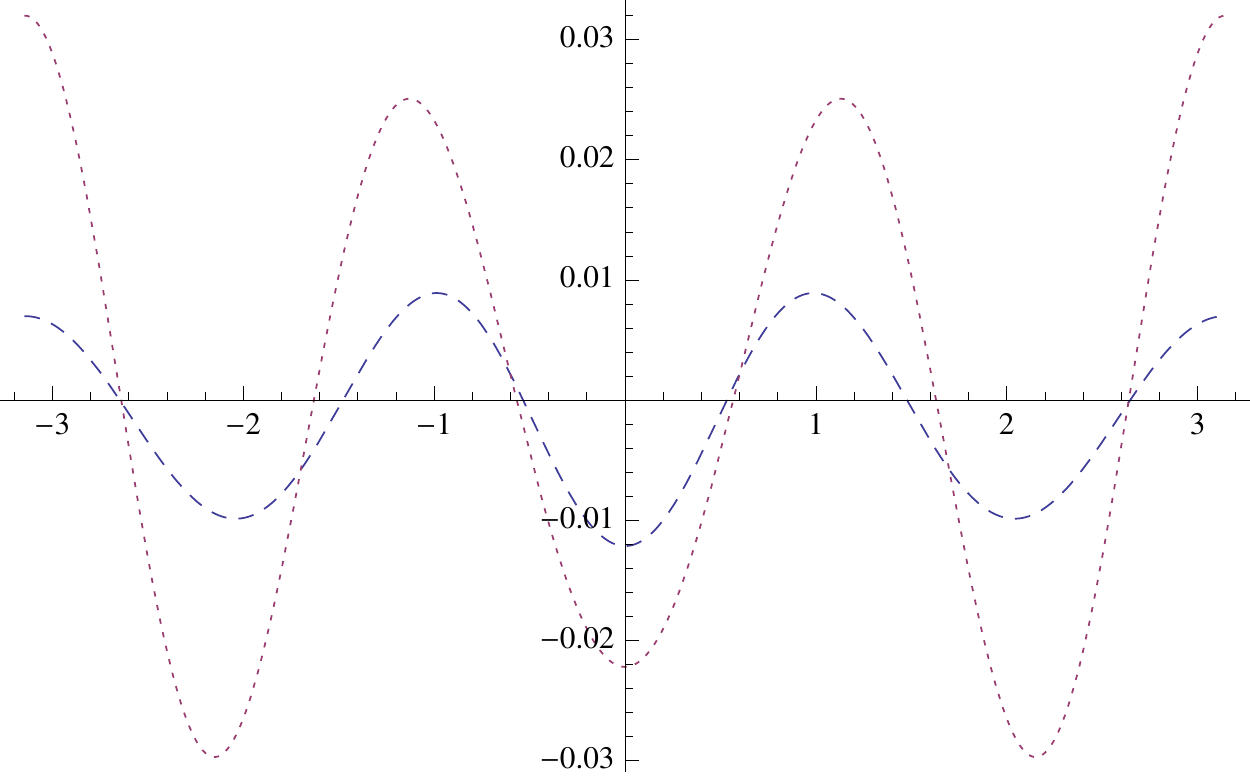}
\end{center}
\caption{The flexural wave amplitude for a cylinder with four control sources (contour plots): $\beta=0.3$, $r_c=1.0$, $N=2$, $\Phi=\pi$. The larger black dot gives the position of the cylinder; the four smaller
black dots give the positions of the control sources: $a=b=u=2.5$. For the upper left figure, $\xi = \pi / 4$, $Q_-=-2.096+1.728\,i$, $Q_+=2.628-1.728\,i$, $P = -3.341$, for the upper right, $\xi = \pi / 3$,  $Q_-=-1.961+1.728\,i$, $Q_+=0.266-1.728\,i$, $P = -2.227$, and for the centre, $\xi = \pi / 2$, $Q_-=-1.404+1.728\,i$, $Q_+=-1.404-1.728\,i$, $P = -1.670$. ($P$ is the amplitude of both sources off the $x$-axis.)
The flexural wave amplitude on a circle of radius 20, as a function of $\theta$,  for a cylinder with four control sources (lower left and right): $\beta=0.3$, $r_c=1.0$, $N=2$, $a=b=u=2.5$, $\xi=\pi/2$, $Q_-=-1.404+1.728\,i$, $Q_+=-1.404-1.728\,i$, $P=-1.670$; blue (dashed): real part, red (dotted): imaginary part. On the lower left, we show the total amplitude, and on the lower right the scattered amplitude. } 
\label{wave_amp_4_sources_piby4and3and2_and_circular_plot_4_sources_piby2}
\end{figure}

\vspace{0cm}
In figure \ref{wave_amp_4_sources_piby4and3and2_and_circular_plot_4_sources_piby2}, we present the flexural wave amplitude pattern for four control sources, two on the $x$-axis as in section \ref{Num_Egs_Two_Sources}, positioned at $(-a,0)$ and $(b,0)$ with amplitudes $Q_-$ and $Q_+$, respectively, and two additional sources symmetrically located off the $x$-axis at $(u\cos \xi, \mp u \sin \xi)$ with the same intensity $P$ (here, $a=b=u=2.5$ and $\xi=\pi/4$ (left), $\xi=\pi/3$ (centre), $\xi = \pi/2$ (right)), thus annulling quadrupole, as well as monopole and dipole outgoing wave amplitudes.  A marked improvement of figure \ref{L_wave_amp_no_sources_R_wave_amp_2_sources} (right) is clearly visible, almost reconstructing the plane wave behind the cylinder.  Whereas there is very little to distinguish between the amplitude fields presented in figure \ref{wave_amp_4_sources_piby4and3and2_and_circular_plot_4_sources_piby2} (upper left and upper right), a visible improvement can be obtained by locating the two additional control sources on the $y-$axis ($\xi=\pi/2$), as shown in the centre contour plot of figure \ref{wave_amp_4_sources_piby4and3and2_and_circular_plot_4_sources_piby2}.

The efficiency of the method can be once again illustrated by the presence of the $\cos 3 \theta$ like term as the leading contributor to the wave amplitude in the lower right-hand plot of figure 
\ref{wave_amp_4_sources_piby4and3and2_and_circular_plot_4_sources_piby2} (compare with bottom right of figure \ref{circular_plot_no_sources_and_two_sources} where $\cos 2 \theta$ is the leading contributor). Note here, that we have only included the total and scattered wave amplitudes for the case of $\xi = \pi/2$.  We do see the presence of $\cos 3 \theta$ in the other two cases (for $\xi = \pi/4, \pi/3$), however, the maximum of the scattered wave amplitudes are in the region of $0.06$ (for $\xi = \pi/4$) and  $0.04$  (for $\xi = \pi/3$).  These maxima are larger than that observed for the case when $\xi = \pi/2$, which is in the region of $0.03$.

If any further improvement to figure \ref{wave_amp_4_sources_piby4and3and2_and_circular_plot_4_sources_piby2} (centre) is required,  it is straightforward to add two more control sources to achieve effectively perfect cloaking. The new two control sources, symmetrically located about the $x$-axis at $(v\cos \eta,\pm v \sin \eta), \,\pi/2< \eta<\pi$, are both assumed to have intensity $R$. Then similar to equation (\ref{P_Greens_fn}), we can write  
\begin{equation}
R \,G(x-v\cos \eta,y \mp v\sin \eta) =
-\frac{R}{8 \beta^2}\sum_{l=-\infty}^\infty  
e^{i l(\theta\mp\eta)}[i H_l^{(1)}(\beta v)J_l(\beta r)-\frac{2}{\pi} K_l(\beta v)I_l(\beta r)]. \label{R_Greens_fn}
\end{equation}

Thus the $n$th order coefficients $A_n, \, B_n$ for the total wave incident on the cylinder (see equation (\ref{An_Bn_4sources})) will have two additional terms $-iRH_n^{(1)}(\beta v) \cos (n\eta)/(4\beta^2)$ and $RK_n(\beta v) \cos (n\eta)/(2\pi\beta^2)$, respectively, and just as before, with $r>\max{(a, b, u, v)}$, we obtain the equations
\begin{equation}
{\cal S}_0[1,1] = {\cal N}_0, \quad i{\cal S}_1[1,1] = {\cal N}_1, \quad -{\cal S}_2[1,1] = {\cal N}_2, \quad -i {\cal S}_3[1,1] = {\cal N}_3, \\
\label{eqns_for_Q+Q-PR}
\end{equation} 
$$
{\cal N}_k = {\cal M}_k + \frac{i R}{4\beta^2} \left[{\cal S}_k[1,1]  H_k^{(1)}(\beta v)+\frac{2 i}{\pi} {\cal S}_k[1,2] K_k(\beta v)+J_k(\beta v)\right]\cos (k{ \eta}), \quad k=0,1,2,3.
$$
Here ${\cal M}_3$ is defined by formula (\ref{M_k}) for $k=3$.

\vspace{-0.1cm}

The intensities $Q_\pm, P, R$ are found by solving the system (\ref{eqns_for_Q+Q-PR}) and the representation for the flexural wave outside the scatterer can be modified accordingly (with the addition of two terms to formula (\ref{Wtotal_add2})).
 
\begin{figure}[H]
\begin{center}
\includegraphics[width=6.5cm]{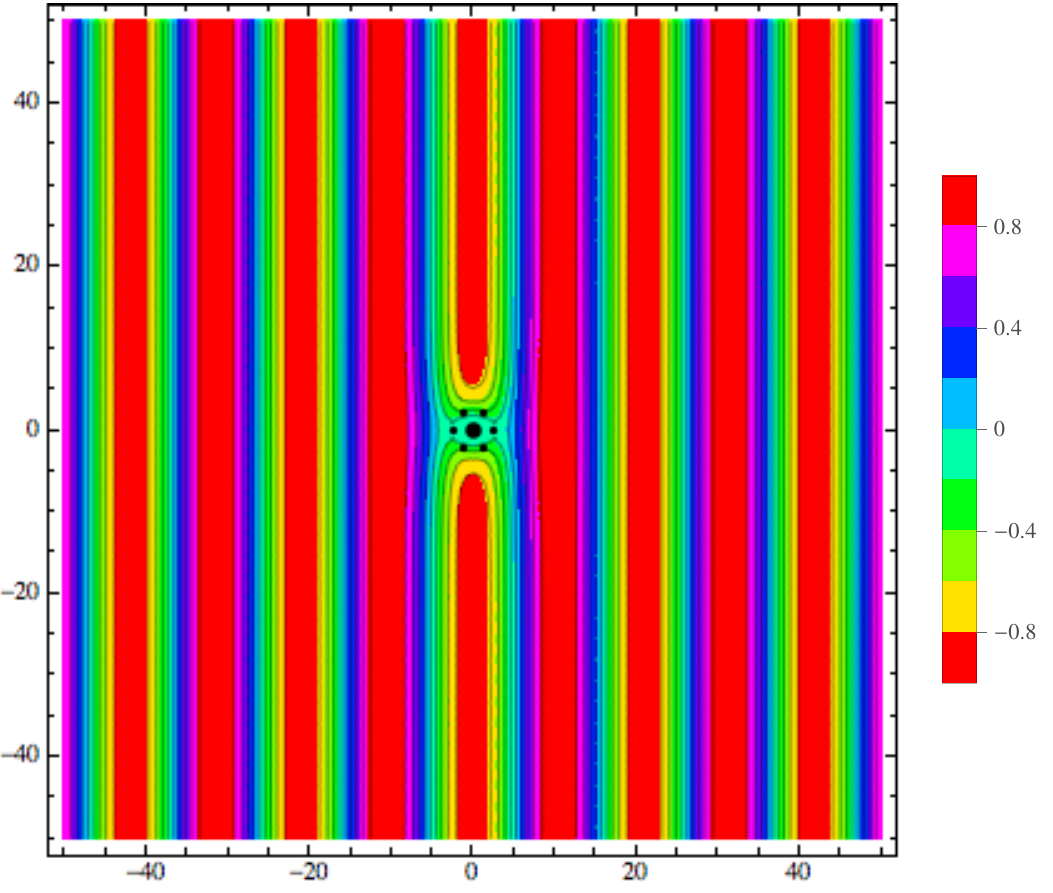} 
\end{center}
\caption{The flexural wave amplitude for a cylinder with six control sources: $\beta=0.3$, $r_c=1.0$, $N=2$, $\Phi=\pi$. The larger black dot gives the position of the cylinder; the six smaller
black dots depict the positions of the control sources: $a=b=u=v=2.5$, $\xi = \pi / 3$, $\eta=2\pi/3$,  $Q_-=-0.848+ 1.146\,i$, $Q_+=-0.848- 1.146\,i$, $P = -1.114- 0.582\,i$, $R=-1.114+ 0.582\,i$ ($P, R$ are the amplitudes of both sources off the $x$-axis in the Ist, IVth and IInd, IIIrd quadrants, respectively.)} 
\label{wave_amp_6_sources}
\end{figure}

Figure \ref{wave_amp_6_sources} presents the scattered wave amplitudes for a cylinder with six control sources, two on the $x$-axis as before, and four off the $x$-axis with $u=v=2.5$ and $\xi = \pi / 3$, $\eta=2\pi/3$. Here, $Q_-=-0.848+ 1.146\,i$, $Q_+=-0.848- 1.146\,i$, $P = -1.114- 0.582\,i$, $R=-1.114+ 0.582\,i$ ($P, R$ are the amplitudes of both sources off the $x$-axis in the Ist, IVth and IInd, IIIrd quadrants, respectively). It is clear that the presence of six control sources improves cloaking from the already effective cloaking in figure \ref{wave_amp_4_sources_piby4and3and2_and_circular_plot_4_sources_piby2} (centre).
 
 \vspace{-0.2cm}
 
\begin{figure}[H]
\begin{center}
\includegraphics[width=6cm]{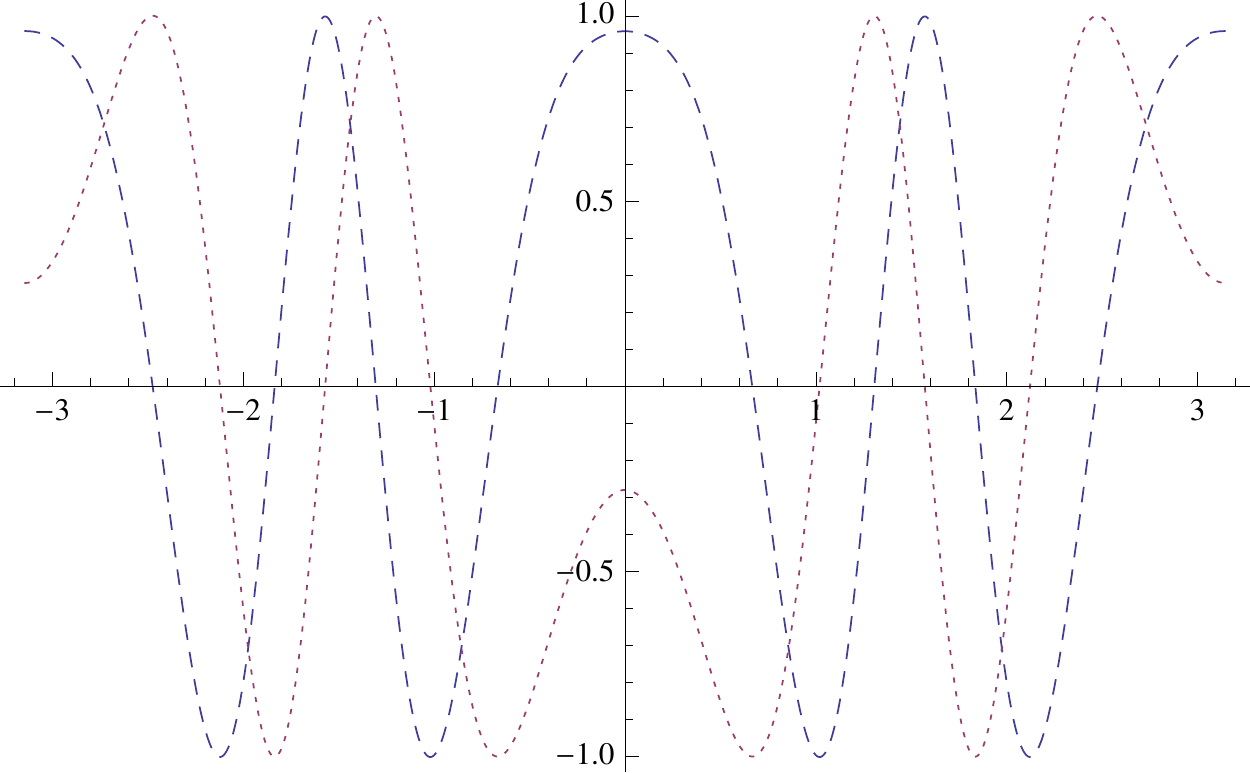}~~
\includegraphics[width=6cm]{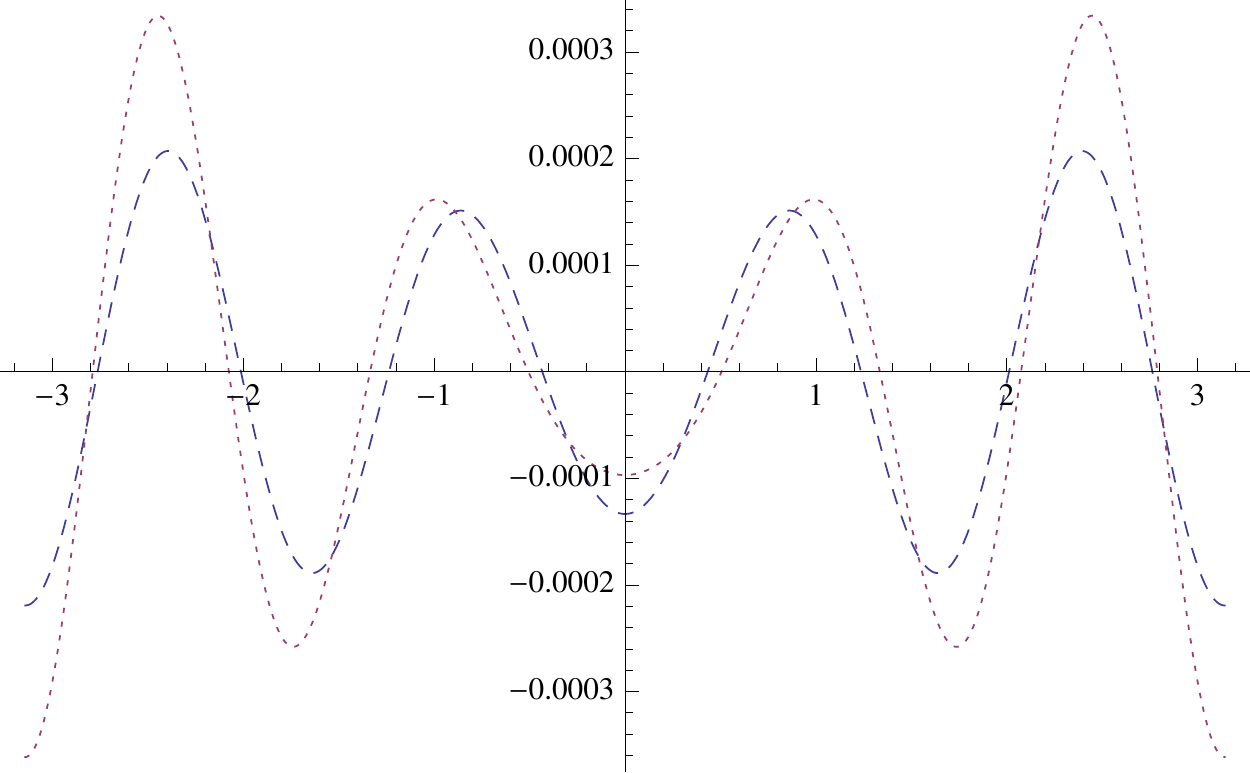}
\end{center}
\caption{The flexural wave amplitude on a circle of radius 40, as a function of $\theta$,  for a cylinder with six control sources: $\beta=0.3$, $r_c=1.0$, $N=2$, $a=b=u=2.5$, $\xi=\pi/3,\,\eta=2\pi/3$,  $Q_-=-0.848+ 1.146\,i$, $Q_+=-0.848- 1.146\,i$, $P = -1.114- 0.582\,i$, $R=-1.114+ 0.582\,i$;
 blue (dashed): real part, red (dotted): imaginary part. On the left, we
show the total amplitude, and on the right the scattered amplitude. } 
\label{circular_plot_6_sources}
\end{figure}

\vspace{-0.2cm}
Another proof of improved cloaking is clearly visible in figure \ref{circular_plot_6_sources} (right) where the leading contribution to the wave amplitude shows $\cos 4\theta$ like behaviour. Note that the scattered amplitude
now peaks close to 0.0003, to be compared with 0.03 in the case of four control sources.

To compare the cloaking achieved using two, four or six control sources, we present the coefficients of $H_n^{(1)} (\beta r)$ and $K_n (\beta r)$ terms 
in table \ref{HnKn_coeffs}.  Note here that for four 
sources, we show coefficients for the configuration where sources away from the $x$-axis are located on the $y$-axis.
We can clearly see from the table that the shaded entries are satisfactorily small for $\tilde{E}_n$ compared to those in the no sources column, meaning that we have successfully eliminated these coefficients of $H_n^{(1)} (\beta r)$ for those particular values of $n$. As payoff for these eliminations, we expect to see an increase in the $\tilde{F}_n$ values. This is true for higher order coefficients ($|n| = 2,3,4$) when we first introduce two sources, however, comparing four and six sources to two, these coefficients remain fairly constant for orders of $n$ that have not eliminated in $\tilde{E}_n$. The fact that there is little change is due to the circular geometry of the inclusion; we do not see any coupling between different orders of the multipole coefficients (which can be seen in equations (\ref{eqns_for_Q+Q-}), (\ref{eqns_for_Q+Q-P}), and (\ref{eqns_for_Q+Q-PR})) for sources placed in a way that does not break the symmetry of the problem.

\vspace{-0cm}

\begin{table}[H]
\begin{center}
\resizebox{13.5cm}{!}{
\begin{tabular}{|c||r|c|c|c|c|} \hline
& $n$ & no sources & 2 sources & 4 sources ($\pi/2$) & 6 sources \\ \hline
\multirow{7}{*}{$\tilde{E}_n$} 
& $-4$ 
& $-2.17\times10^{-12}-1.47 \times10^{-6} \,i$ 
& $9.94\times10^{-9} + 0.007  \,i$
& $9.94\times10^{-9} + 0.007  \,i$
& $-8.62\times10^{-10} - 0.0006\,i$
 \\ \cline{2-6}
& $-3$ 
& $0.0004- 1.53\times10^{-7} \,i$ 
&$0.04 - 0.00001 \,i$ 
& $0.04 - 0.00001 \,i$ 
& \cellcolor{gray!25}$-1.20\times10^{-17} - 1.64\times10^{-17}\, i$
\\ \cline{2-6}
& $-2$ 
&$0.001 + 0.03 \,i$
& $0.02 + 0.43 \, i$
& \cellcolor{gray!25}$1.18\times10^{-16} - 4.16\times10^{-17}\, i$
&  \cellcolor{gray!25}$-5.36\times10^{-17} + 1.39\times10^{-17}  \,i$
\\ \cline{2-6}
& $-1$ 
&$-0.43 + 0.24 \,i$ 
& \cellcolor{gray!25}$-7.47\times10^{-7} + 4.24\times10^{-7} \, i$
& \cellcolor{gray!25}$-2.22\times10^{-16} - 1.85\times10^{-17}  \, i$
&  \cellcolor{gray!25}$2.22\times10^{-16} -1.88\times10^{-16} \,i$
\\ \cline{2-6}
& $0$ 
& $-0.94 + 0.23 \, i$ 
& \cellcolor{gray!25}$-4.72\times10^{-7} + 1.15\times10^{-7}\, i$
& \cellcolor{gray!25}$-5.50\times10^{-16}$
& \cellcolor{gray!25}$-1.36\times10^{-16} + 1.78\times10^{-15} \,i$  
\\ \cline{2-6}
& $1$ 
&  $0.43 - 0.24 \, i$
& \cellcolor{gray!25}$7.47\times10^{-7} - 4.24\times10^{-7} \, i$
& \cellcolor{gray!25}$2.22\times10^{-16} + 1.85\times10^{-17}  \, i$
& \cellcolor{gray!25}$-6.66\times10^{-16} - 1.63\times10^{-16} \,i$
\\ \cline{2-6}
& $2$ 
&$0.001 + 0.03 \,i$ 
& $0.02 + 0.43 \, i$
& \cellcolor{gray!25}$1.18\times10^{-16} - 4.16\times10^{-17}\, i$
& \cellcolor{gray!25}$6.05\times10^{-18} - 2.78\times10^{-17}  \,i$
\\ \cline{2-6}
& $3$ 
& $-0.0004 + 1.53\times10^{-7} \,i$ 
& $-0.04 + 0.00001 \,i$ 
& $-0.04 + 0.00001 \,i$ 
&  \cellcolor{gray!25}$1.20\times10^{-17} + 1.64\times10^{-17}  \,i$ 
\\ \cline{2-6}
& $4$ 
& $-2.17\times10^{-12}-1.47 \times10^{-6} \,i$
& $9.94\times10^{-9} + 0.007  \,i$
& $9.94\times10^{-9} + 0.007  \,i$
& $-8.62\times10^{-10} - 0.0006\,i$
\\ \hline\hline
\multirow{7}{*}{$\tilde{F}_n$} 
& $-4$ 
& $9.49\times10^{-7} - 1.40\times10^{-12} \,i$  
& $-0.005 + 6.40\times10^{-9}  \, i$
& $-0.005 + 6.40\times10^{-9} \, i$
& $0.0004 - 5.55\times10^{-10}  \,i$
\\ \cline{2-6}
& $-3$ 
& $-9.90\times10^{-8} - 0.0003 \,i$ 
& $-9.67\times10^{-6} - 0.03 \,i$
& $-9.67\times10^{-6} - 0.03 \,i$
& \cellcolor{gray!25}$-1.28\times10^{-17} + 1.55\times10^{-5} \,i$
\\ \cline{2-6}
& $-2$ 
&$-0.02 + 0.0008 \,i$
& $-0.31 + 0.01\,i$
& \cellcolor{gray!25}$0.002 +8.52\times10^{-17} i$
& \cellcolor{gray!25}$0.002 - 3.49\times10^{-17}  \,i$
\\ \cline{2-6}
& $-1$ 
&$0.16 + 0.29 \,i $
& \cellcolor{gray!25}$2.82\times10^{-7} - 0.12 \,i$
& \cellcolor{gray!25}$3.86\times10^{-17} - 0.12 \,i$
& \cellcolor{gray!25}$-1.54\times10^{-16} - 0.12  \,i$
\\ \cline{2-6}
& $0$ 
&$-0.18 - 0.72 \,i$
& \cellcolor{gray!25}$-0.96 - 3.60\times10^{-7} \,i$
& \cellcolor{gray!25}$-0.96 - 1.16\times10^{-16} \,i$
& \cellcolor{gray!25}$-0.96 - 6.92\times10^{-16}  \,i$ 
\\ \cline{2-6}
& $1$ 
&$0.16 + 0.29 \,i $ 
& \cellcolor{gray!25}$2.82\times10^{-7} - 0.12 \,i$
& \cellcolor{gray!25}$3.86\times10^{-17} - 0.12 \,i$
&  \cellcolor{gray!25}$3.93\times10^{-16} - 0.12 \,i$
\\ \cline{2-6}
& $2$ 
& $-0.02 + 0.0008 \,i$
& $-0.31 + 0.01\,i$
& \cellcolor{gray!25}$0.002 +8.52\times10^{-17} i$
& \cellcolor{gray!25} $0.002 - 8.67\times10^{-18}  \,i$
\\ \cline{2-6}
& $3$ 
& $-9.90\times10^{-8} - 0.0003 \,i$  
&  $-9.67\times10^{-6} - 0.03 \, i$
& $-9.67\times10^{-6} - 0.03 \,i$
&  \cellcolor{gray!25}$-1.28\times10^{-17} + 1.55\times10^{-5} \,i$
\\ \cline{2-6}
& $4$ 
&$9.49\times10^{-7} - 1.40\times10^{-12} \,i$  
& $-0.005 + 6.40\times10^{-9}  \, i$
& $-0.005 + 6.40\times10^{-9} \, i$
& $0.0004 - 5.55\times10^{-10}  \,i$
\\ \hline 
\end{tabular} 
}
\caption{$\tilde{E}_n$ and $\tilde{F}_n$ denote the coefficients of $H_n^{(1)} (\beta r)$ and $K_n (\beta r)$ terms, respectively (see  (\ref{Hn_coeff}) and (\ref{Kn_coeff}) for a configuration of two control sources located on the $x$-axis). Shaded entries correspond to the values of $n\,(=0,\pm1,\dots,\pm4)$ for which we have eliminated the $H_n (\beta r)$ coefficients.}
\label{HnKn_coeffs}
\end{center}
\end{table}

\vspace{-0.5cm}

\begin{figure}[H]
\begin{center}
\includegraphics[width=6.5cm]{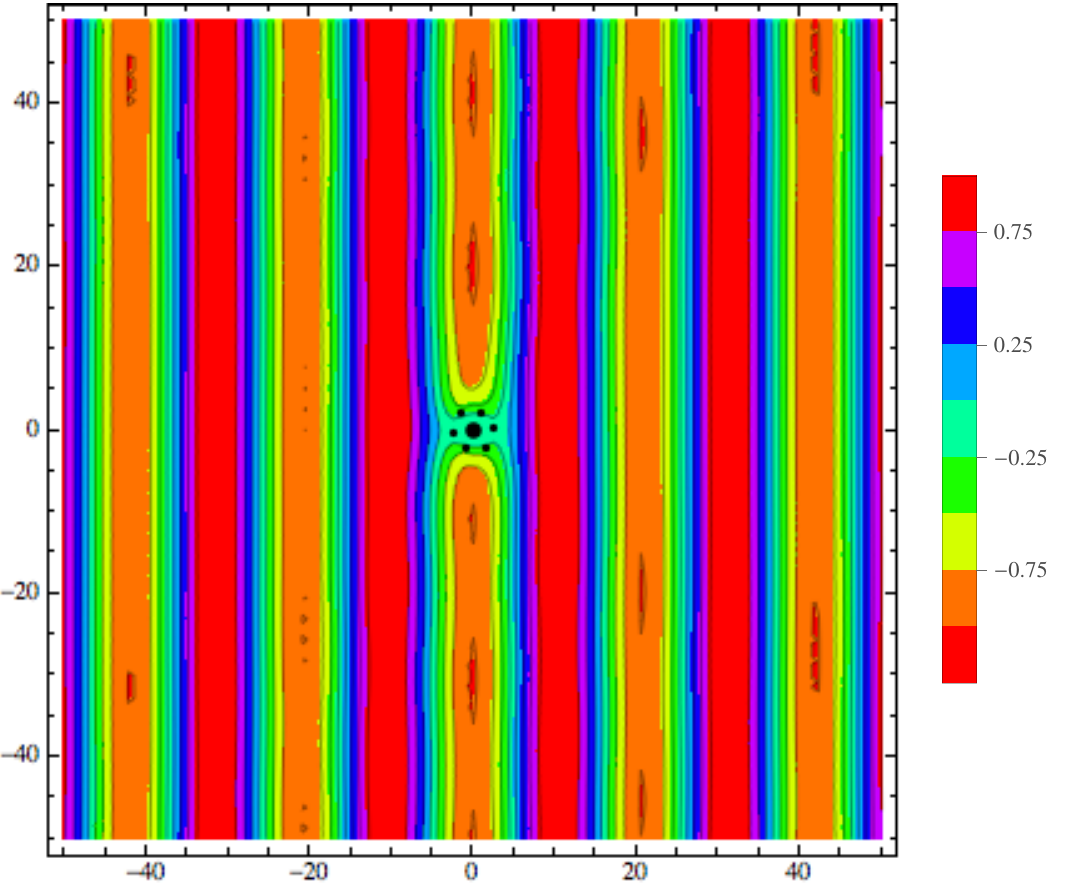}~~
\includegraphics[width=6.5cm]{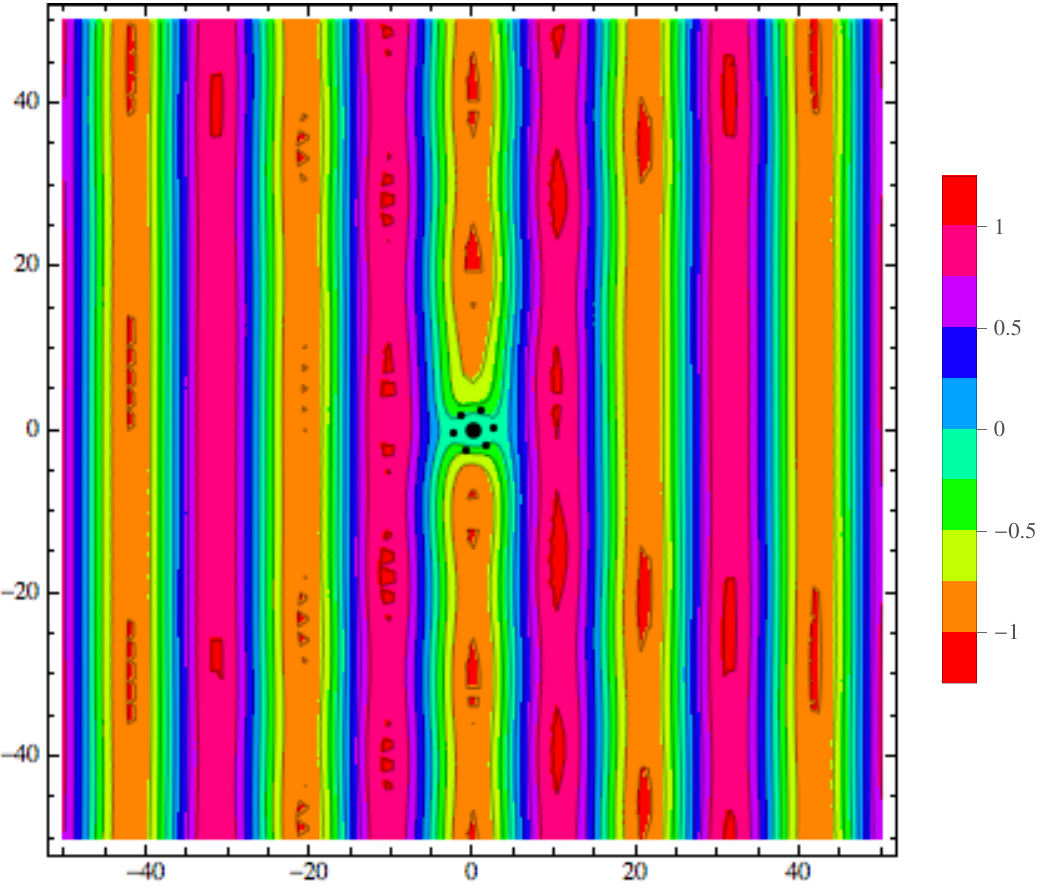} \\
\includegraphics[width=6.5cm]{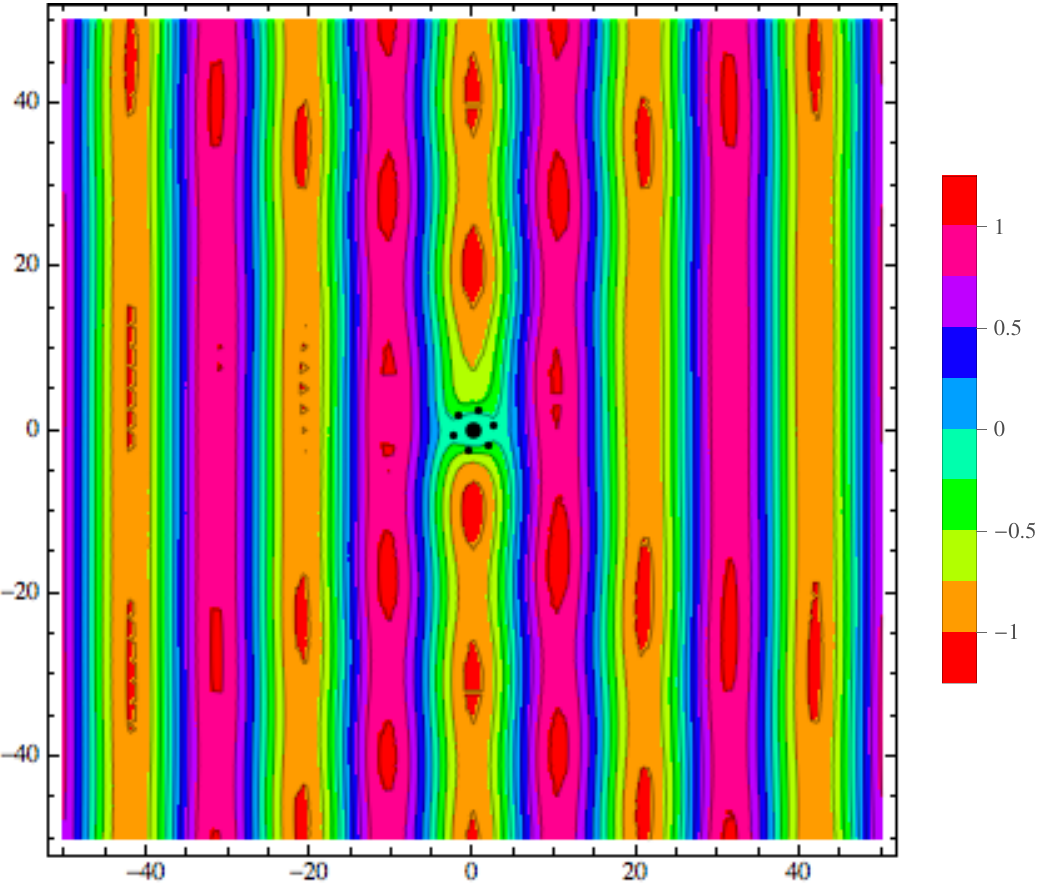}~~
\includegraphics[width=6.5cm]{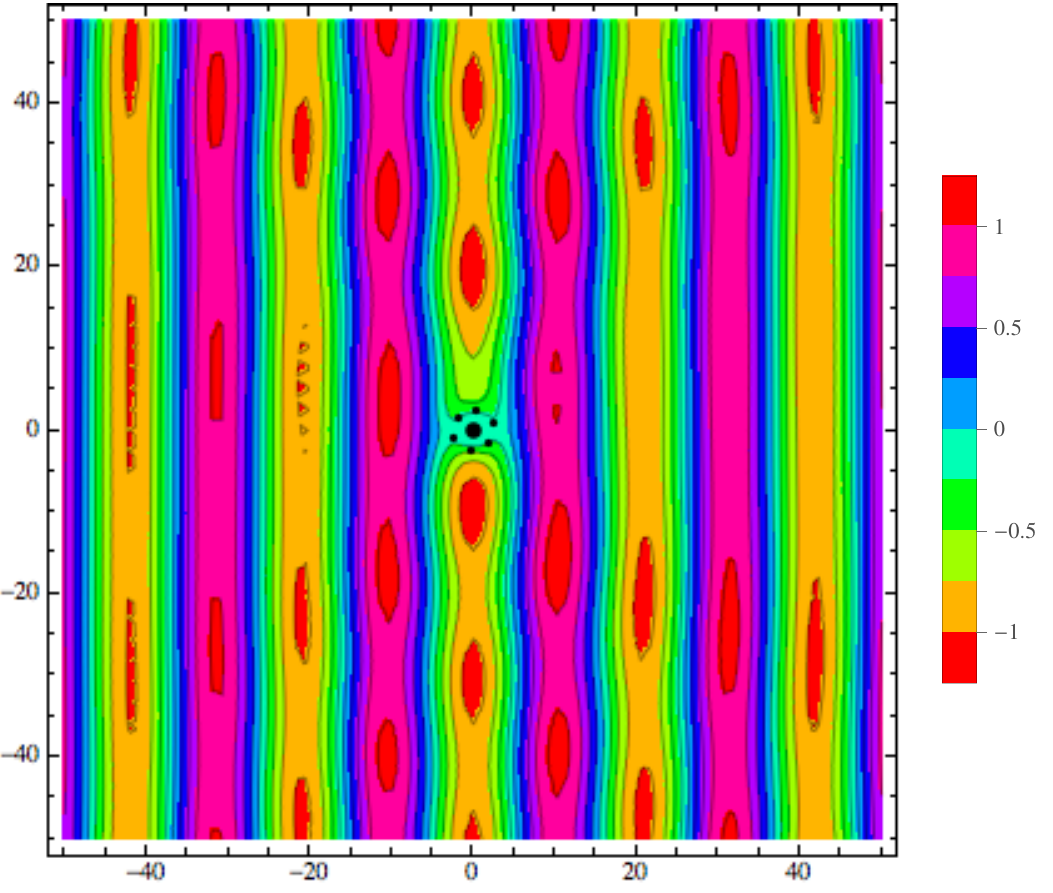}
\end{center}
\caption{The flexural wave amplitude for a cylinder with six control sources: $\beta=0.3$, $r_c=1.0$, $N=2$, $\Phi=\pi$. The configuration of six control sources in Figure \ref{wave_amp_6_sources}  are rotated by an angle of $5$ degrees (top left), $10$ degrees (top right), $15$ degrees (bottom left), $20$ degrees (bottom right), respectively.} 
\label{wave_amp_6_sources_rot}
\end{figure}

At this point, a natural extension is to investigate the robustness of our active cloaking to rotation of the six control sources. We do this by using the same control source amplitudes and rotating the entire configuration by  an angle in anti-clockwise direction. The results are presented in figure \ref{wave_amp_6_sources_rot}, which indicate that small rotations do not affect the cloaking very much.

\vspace{-0.2cm}

\section{Green's function versus active cloak for a rigid inclusion}
\label{Scatter_CircularWave}

\vspace{-0.4cm}

The concept of active cloaking is, of course, not limited to plane waves; we now consider the problem 
where the incident wave is represented by the Green's function corresponding to a remote point source.
Thus, a cylindrical wave is generated by a point source at  $(-c, 0)$, placed sufficiently far away from the rigid inclusion.
Hence, the incident wave can be represented as
\begin{equation}
w_0(r, \theta)= 
-\frac{1}{8 \beta^2}\sum_{l=-\infty}^\infty  
(-1)^l\left[i  H_l^{(1)}(\beta c)J_l(\beta r)-\frac{2}{\pi} K_l(\beta c)I_l(\beta r)\right]e^{i l \theta}, \label{incident_cylindrical}
\end{equation}
where we assume that the amplitude of the incident wave is unity.

In this particular case, the coefficients of the $n$th order total wave incident on the cylinder now read
{\small 
\begin{equation}
\begin{split}
A_n &= - \frac{i}{8 \beta^2} (-1)^n H_n^{(1)} (\beta c)- i\frac{Q_+}{8\beta^2}H_n^{(1)}(\beta b)-i\frac{Q_-}{8\beta^2}(-1)^n H_n^{(1)}(\beta a)  
 -\frac{i P}{4\beta^2} H_n^{(1)} (\beta u)\, \cos (n\xi) -\frac{iR}{4\beta^2}H_n^{(1)}(\beta v) \cos (n\eta), \\
B_n &= \frac{1}{4 \beta^2 \pi} (-1)^n K_n (\beta c) + \frac{Q_+}{4\pi \beta^2}K_n(\beta b)+\frac{Q_-}{4\pi \beta^2}(-1)^n K_n(\beta a)   +\frac{P}{2\pi\beta^2} K_n(\beta u) \, \cos (n\xi) +\frac{R}{2\pi\beta^2}K_n(\beta v) \cos (n\eta). 
\label{An_Bn_6sources_incident_cylindrical}
\end{split}
\end{equation}
}
We note that formulae (\ref{An_Bn_6sources_incident_cylindrical}) is given for six  control sources around the inclusion ($-3 \leq n \leq 3$); $P=R=0$ if there are only two control sources ($-1 \leq n \leq 1$), and $R=0$ if there are four ($-2 \leq n \leq 2$). 

For $r > \max{(a, b, u,v)}$ but $r< c$, we can obtain the equations to find the amplitudes of the control sources $Q_\pm, P, R$ from equation (\ref{eqns_for_Q+Q-PR}) by replacing the four left-hand sides by 
\begin{equation}\label{cylindrical_wave_lhs}
 - \frac{i}{8 \beta^2} (-1)^n H_n^{(1)} (\beta c) {\cal S}_n[1,1] +  \frac{1}{4 \beta^2 \pi} (-1)^n K_n (\beta c) {\cal S}_n[1,2], \quad n = 0,1,2,3,
\end{equation}
respectively.

\begin{figure}[H]
\begin{center}
\includegraphics[width=6.5cm]{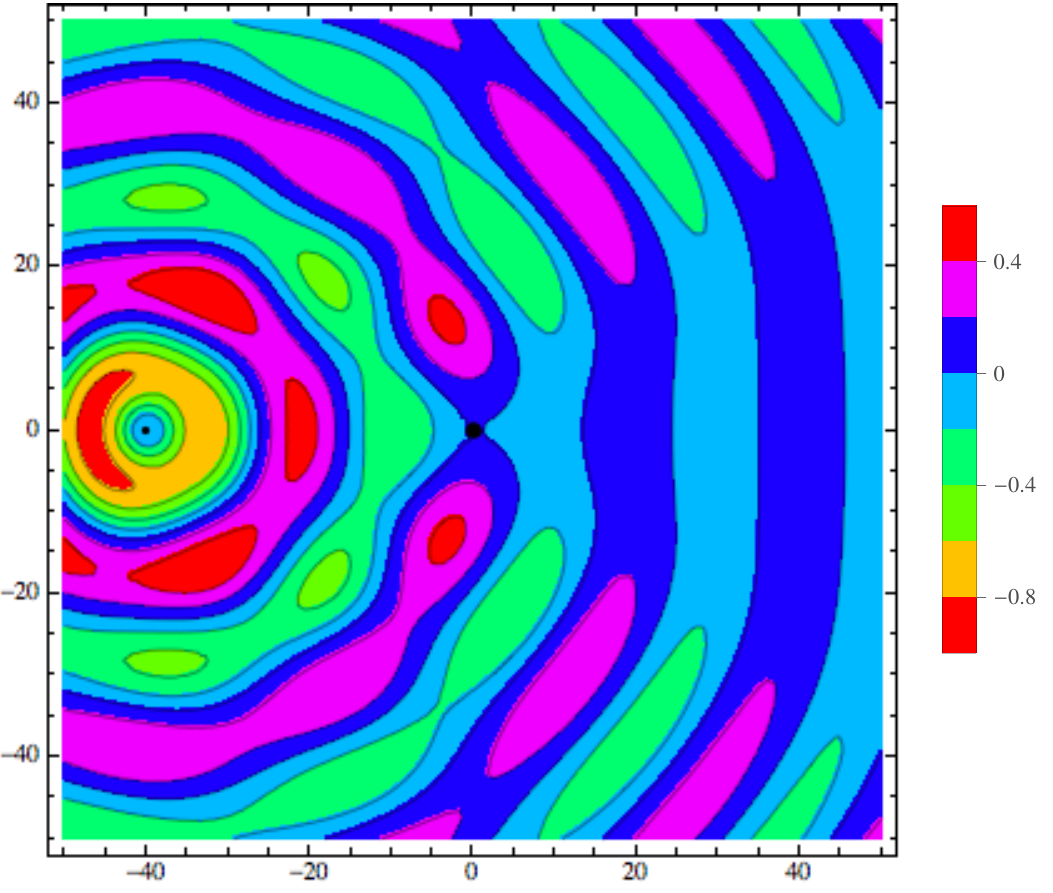}~~
\includegraphics[width=6.5cm]{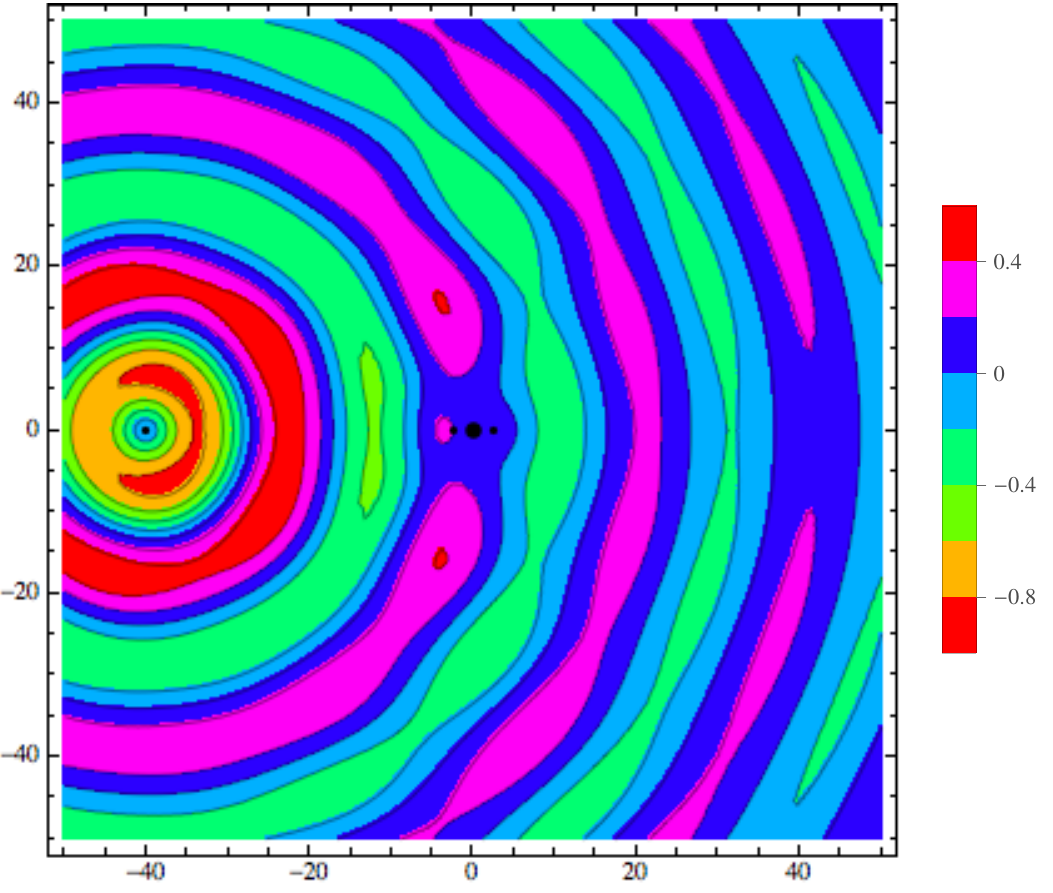}
\end{center}
\caption{The flexural wave amplitude for a cylinder with no control sources (left), and for a cylinder with two control sources (right): $\beta=0.3$, $r_c=1.0$, $N=2$, $\Phi=\pi$ (both), $a=b=2.5$, $Q_-=1.099 -0.333i$ and $Q_+=0.825+0.740i$, (right). The large black dot depicts the position of the cylinder, the small dot the position of the point source emitting the incident wave and the two small dots around the cylinder the control sources (right).} \label{wave_amp_no_sources_incps_and_2_sources_incps}
\end{figure}

Now, making the necessary adjustments to formula (\ref{Wtotal_add2}), we can find the amplitude for the flexural wave outside the scatterer.  In figure \ref{wave_amp_no_sources_incps_and_2_sources_incps}, the left-hand plot depicts the wave intensities for a cylinder with no control sources present, whilst the right-hand plot depicts the amplitude for two control sources positioned along the $x$-axis as in section \ref{Num_Egs_Two_Sources}. The control sources in the right-hand plot have intensities $Q_-=1.099 -0.333i$ and $Q_+=0.825+0.740i$.  
If there are no control sources present, we see a large shadow region behind the cylinder, but it is clear that when the 
pair of sources are added, we begin to reconstruct the cylindrical wave in this region. 

\vspace{-0.3cm}

\vspace{-0.5cm}

\begin{figure}[H]
\begin{center}
\includegraphics[width=6.4cm]{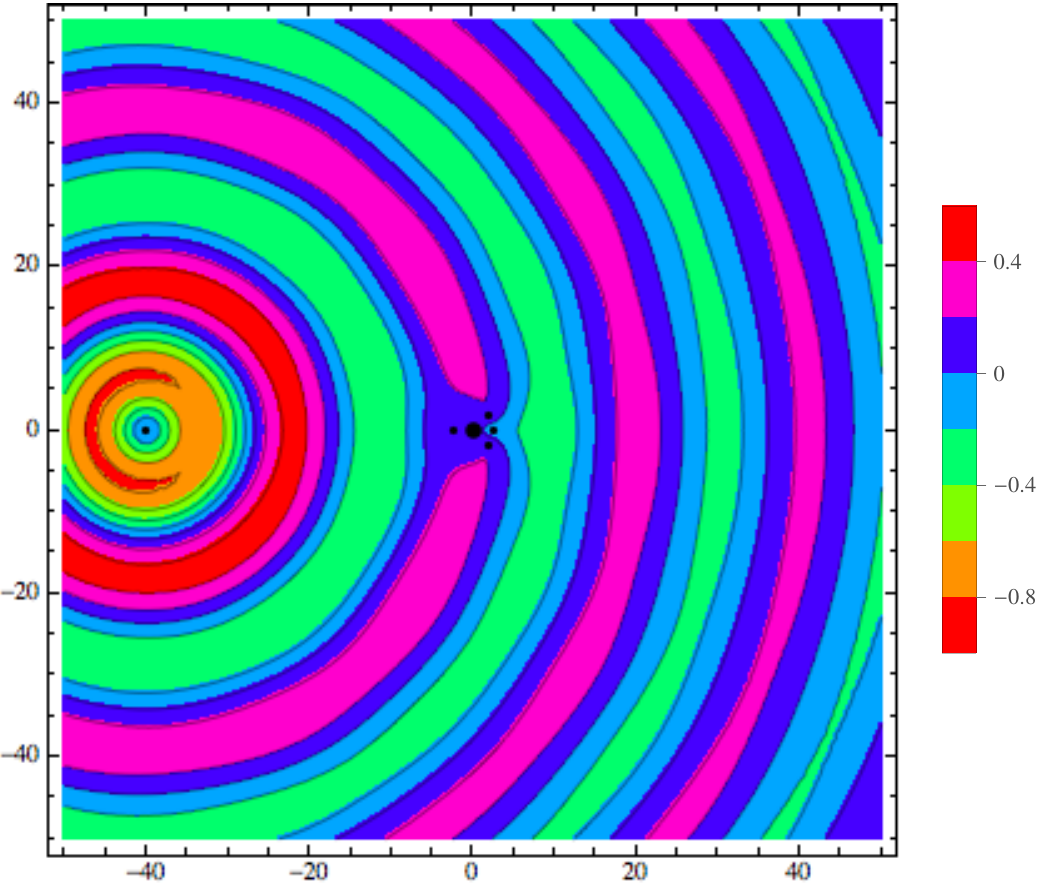}~~\includegraphics[width=6.4cm]{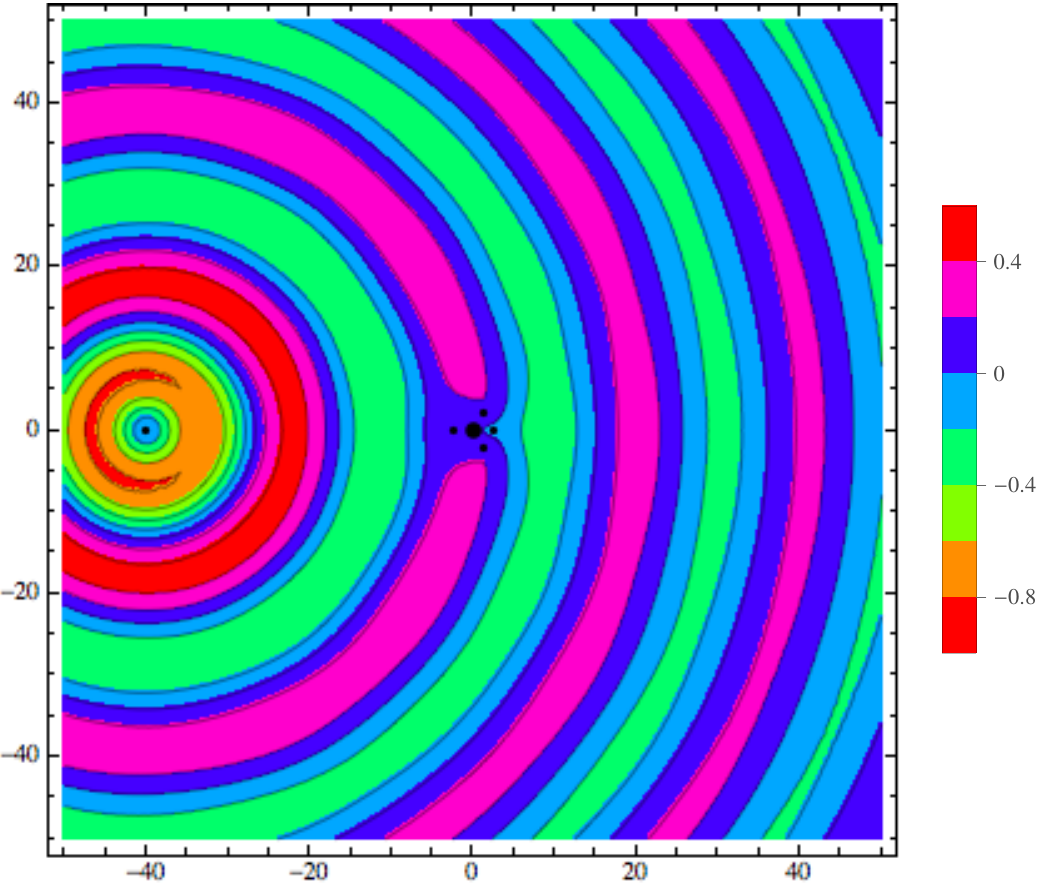}\\
\includegraphics[width=6.4cm]{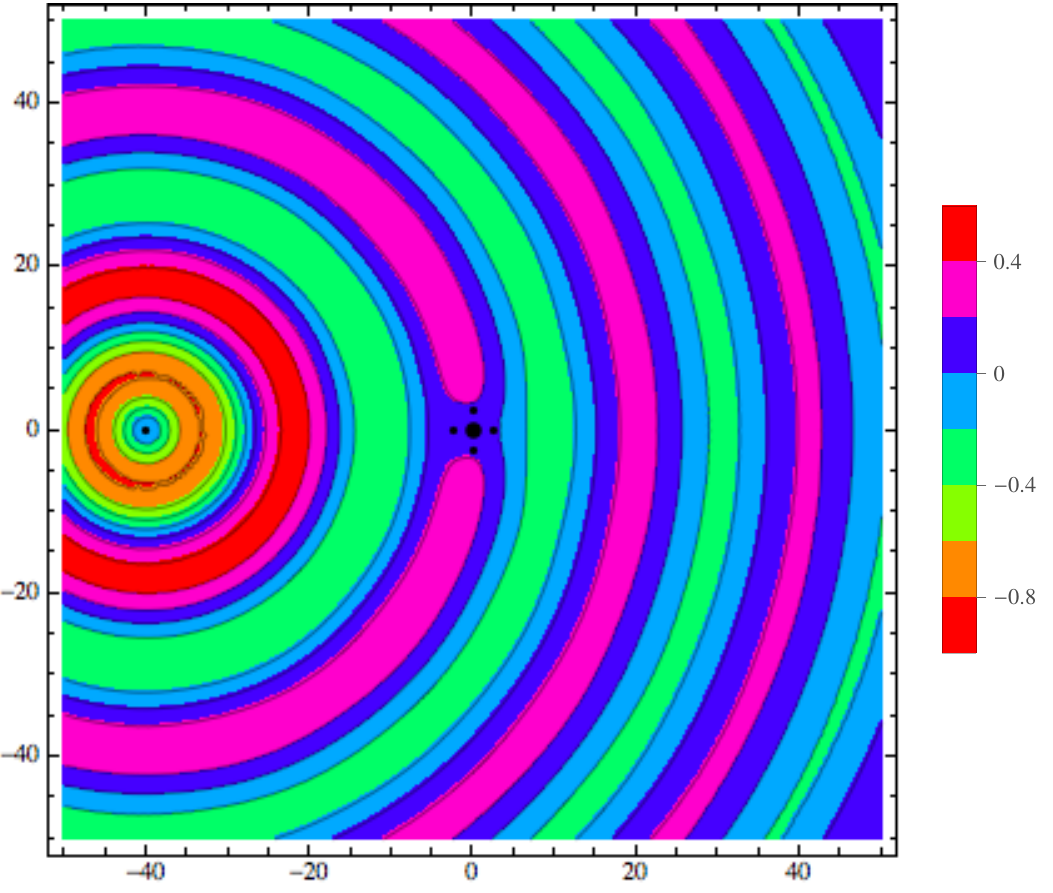}
\end{center}
\caption{
The flexural wave amplitude for a cylinder with four control sources: $\beta=0.3$, $r_c=1.0$, $N=2$, $\Phi=\pi$,  $a=b=u=2.5$ (for all three); $Q_-=0.794-0.402i,\,Q_+=-0.953+0.339i,\, P=1.042+0.235i$ (left), $Q_-=0.752-0.411i,\,Q_+=-0.217+0.505i,\, P=0.694+0.157i$ (centre), $Q_-=0.578-0.450i,\,Q_+=-0.304+0.623i,\, P=0.521+0.118i$ (right). The large black dot depicts the position of the cylinder, the small dot on the far left the position of the point source emitting the incident wave and the four small dots around the cylinder the control sources ($\xi=\pi/4, \,\xi=\pi/3, \,\xi=\pi/2$ from left to right), respectively. ($P$ is the amplitude of both sources off the $x$-axis.)}
\label{wave_amp_4_sources_incps}
\end{figure}

Amplitude plots associated with four control sources around the inclusion with the same configuration as in section \ref{cloaking_additional _control_sources} are presented in figure \ref{wave_amp_4_sources_incps}. The progression of the efficiency of cloaking from $\xi=\pi/4$ or $\xi=\pi/3$ to $\xi=\pi/2$ is clearly visible.

\vspace{-0.2cm}

\begin{figure}[H]
\begin{center}
\includegraphics[width=6.5cm]{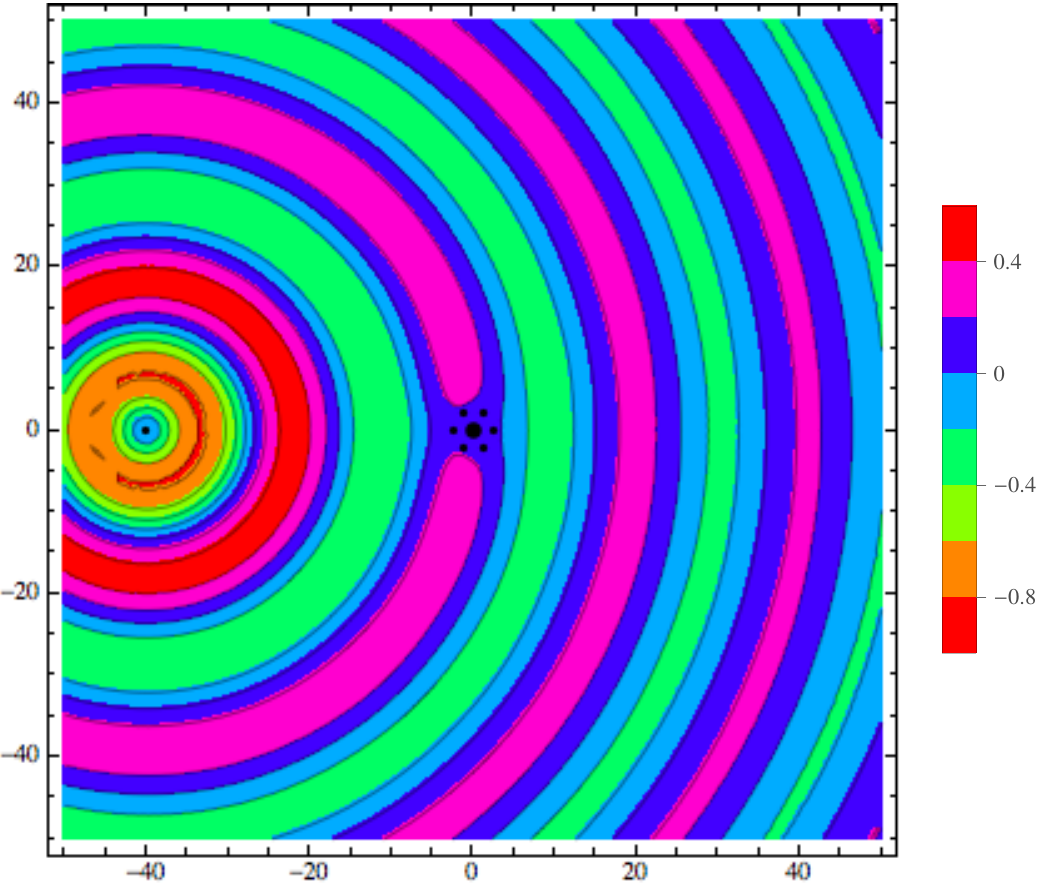}
\end{center}
\caption{The flexural wave amplitude for a cylinder with six control sources: $\beta=0.3$, $r_c=1.0$, $N=2$, $\Phi=\pi$, $a=b=u=2.5$;  $Q_-=0.282-0.292i,\,Q_+=-0.253+0.386i,\, P=0.224+0.275, \, R=0.470 - 0.119 i$. The large black dot depicts the position of the cylinder, the small dot on the far left the position of the point source emitting the incident wave and the six small dots around the cylinder the control sources, respectively. ($P, R$ are the amplitudes of both sources as in figure \ref{wave_amp_6_sources}.)}
\label{wave_amp_6_sources_incps}
\end{figure}

Finally, we illustrate effectively perfect cloaking with six control sources in figure \ref{wave_amp_6_sources_incps}. 

\vspace{-0.2cm}

\section{An arbitrarily shaped scatterer}
\label{arbitrarily_shaped}

The use of active cloaking sources presented in the previous sections can be extended to cloaking objects of arbitrary shape.  To illustrate this we assume that the plate has an arbitrarily shaped hole, in which we locate the origin.

\begin{figure}[H]
\begin{center}
\includegraphics[width=5.5cm]{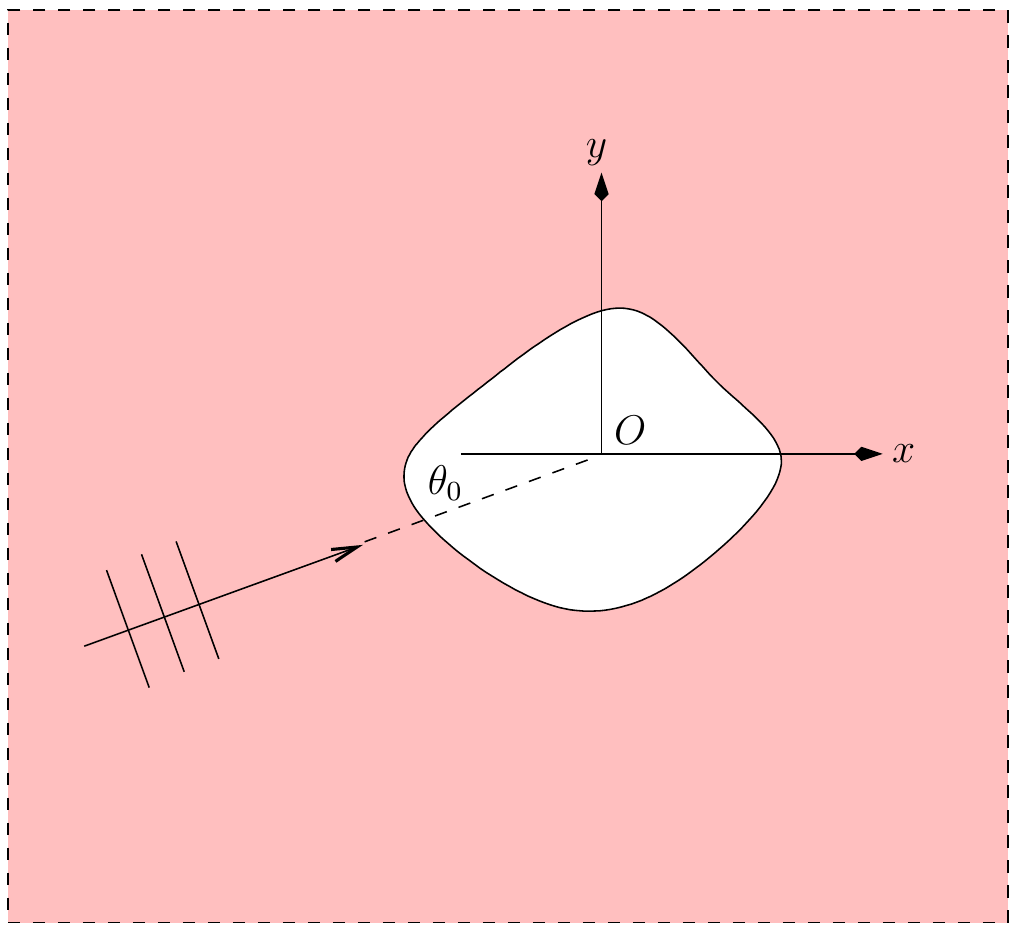} \,\,\,\,\,
\includegraphics[width=5.5cm,  height=5.1cm]{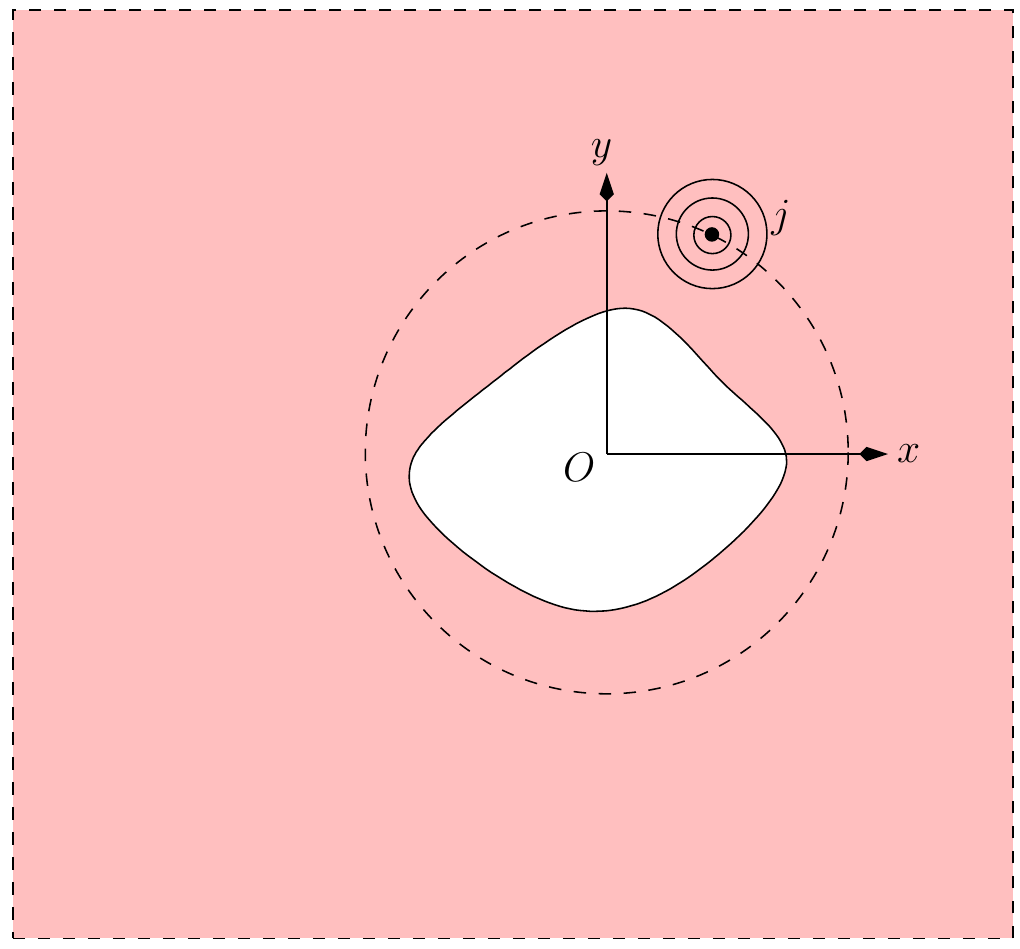}
\end{center}
\caption{Left: Model problem 1: Plane wave scattering by an arbitrarily shaped scatterer. The plane wave is incident at an angle $\theta_0$. Right: Model problem 2: Cylindrical wave scattering by an arbitrarily shaped scatterer. 
} \label{model_probs}
\end{figure}

We thus consider two model problems entirely independent of each other, as illustrated in figure \ref{model_probs}.

{\bf Model problems}

\underline{Plane wave interacting with arbitrarily shaped scatterer.} \\
The displacement field due to the plane wave perturbation, $w^{(p)}$, satisfies the boundary value problem
\begin{equation}
\begin{split}
\Delta^2 w^{(p)}  - \beta^4 w^{(p)} & =0 \quad \mbox{in} \,\,\mathbb{R}^2 \setminus \bar{{\cal D}}, \\
w^{(p)} = \frac{\partial w^{(p)}} {\partial n} &=0 \quad \mbox{on}\,\, \partial {\cal D},
\end{split}
\end{equation}
where ${\cal D}$ is the arbitrarily shaped scatterer. We note that  $w^{(p)}$ can be written as the sum of the incident field $w^{(i)}$ and the scattered field $w_{sc}^{(p)}$, that is
\begin{equation}\label{plane_wave_field}
w^{(p)} ({\bf x}) = w^{(i)} ({\bf x}) +w_{sc}^{(p)} ({\bf x}),
\end{equation}
where $w^{(i)}$ is the same as $w_0$ defined in (\ref{incident_plane_wave}) and (\ref{incident_plane_wave_polar_form}).
The scattered field $w_{sc}^{(p)}$ (see, for example, the sum in formula (\ref{Wtotal})) has the asymptotic representation 
\begin{equation}\label{plane_wave_scattered_field}
w_{sc}^{(p)}  = \sum_{n=-\infty}^\infty \left[ {E}_n^{(p)} H_{n}^{(1)}(\beta r) + {F}_n^{(p)} K_{n}(\beta r) \right]
\, e^{i n (\theta-\theta_0)}
\sim 
\sum_{n=-\infty}^\infty {E}_n^{(p)} e^{-in\theta_0} H_{n}^{(1)}(\beta r) e^{in\theta},
\end{equation}
where ${E}_n^{(p)}$ are constant coefficients, 
since $K_n(\beta r) = {\it O} (\exp(-\beta r)/(\beta r))$ as $\beta r \gg 1$ (see formula 9.7.2 in \cite{MA_IAS}).

\underline{Cylindrical wave interacting with arbitrarily shaped scatterer.} \\The displacement field due to the cylindrical wave perturbation $w^{(s,j)}$, emitting from source $j$, satisfies the boundary value problem
\begin{equation}
\begin{split}
\Delta^2 w^{(s,j)}  - \beta^4 w^{(s,j)} + \delta({\bf x} - {\bf X}^{(j)})& =0 \quad \mbox{in} \,\,\mathbb{R}^2 \setminus \bar{{\cal D}}, \\
w^{(s,j)} = \frac{\partial w^{(s,j)}} {\partial n} &=0 \quad \mbox{on}\,\, \partial {\cal D},
\end{split}
\end{equation}
where $\delta({\bf x} - {\bf X}^{(j)})$ denotes the Dirac delta function, centred at ${\bf X}^{(j)}$. In fact, $w^{(s,j)}$admits the solution
\begin{equation}\label{arbitrary_source_field}
\begin{split}
w^{(s,j)} ({\bf x}) &= G({\bf x} - {\bf X}^{(j)} ) + w_{sc}^{(s,j)} ({\bf x}), \\
&\sim \sum_{n=-\infty}^\infty {\cal A}_n^{(s,j)} H_{n}^{(1)}(\beta r) e^{in\theta},  \quad j=\,1, \dots,N,
\end{split}
\end{equation}
where exponentially small terms, that satisfy the modified Helmholtz equation, are not shown.   
Here $G({\bf x} - {\bf X}^{(j)} )$ denotes the Green's function for the biharmonic operator, $w_{sc}^{(s,j)}$ is the scattered field due to the unit source at ${\bf X}^{(j)}$, and ${\cal A}_n^{(s,j)}$ are constant coefficients.

Assuming that the coefficients ${E}_n^{(p)}$ and ${\cal A}_n^{(s,j)}$ are given,  an active cloaking is achieved by introducing a set of $N$ control sources of complex intensities ${\cal Q}_j $ placed at the points ${\bf X}^{(j)}$ around the scatterer ${\cal D}$.  
After the truncation to order $K$ in the expansions (\ref{plane_wave_scattered_field}), (\ref{arbitrary_source_field}), we choose $N=2 K+1$, so that    the total displacement field $w^{total}$ is approximately equal to the incident field $w^{(i)}$, that is
\begin{equation}\label{total_field}
w^{total} = w^{(p)} + \sum_{j=1}^N {\cal Q}_j w^{(s,j)} \approx w^{(i)}.
\end{equation}
To find ${\cal Q}_j$, we substitute (\ref{plane_wave_scattered_field})  and (\ref{arbitrary_source_field}) into (\ref{total_field}), and obtain the following system of linear algebraic equations
\begin{equation}\label{fourier_coeff_eqn}
{E}_k^{(p)} e^{-ik \theta_0} + \sum_{j=1}^{2K+1} {\cal Q}_j {\cal A}_k^{(s,j)} = 0,
 \quad k = -K,\dots, K.
 \end{equation}
 
{\bf Numerical computations}

\begin{figure}[H]
\begin{center}
\includegraphics[width=7cm]{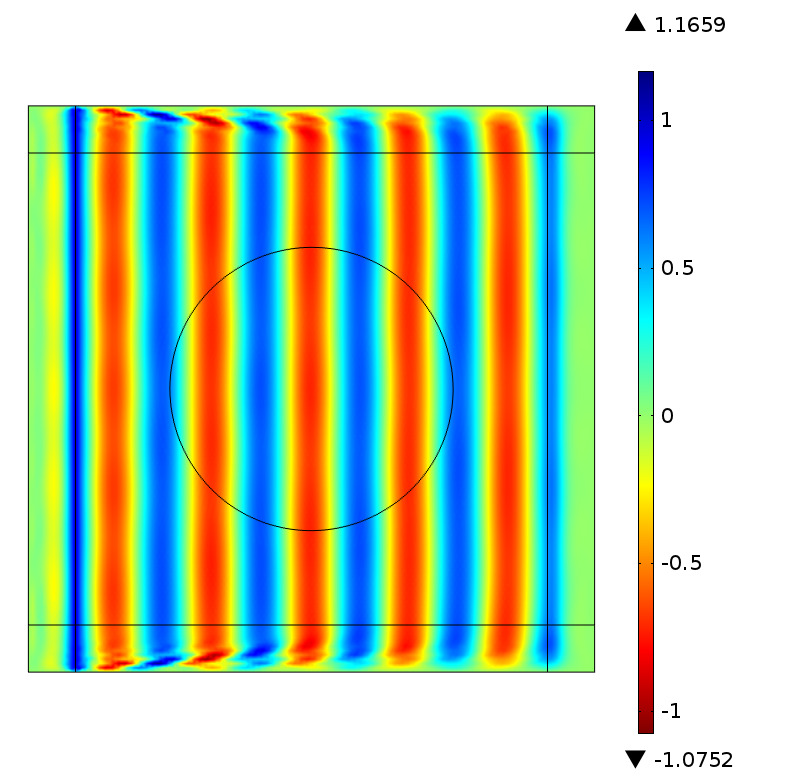}
\includegraphics[width=7cm]{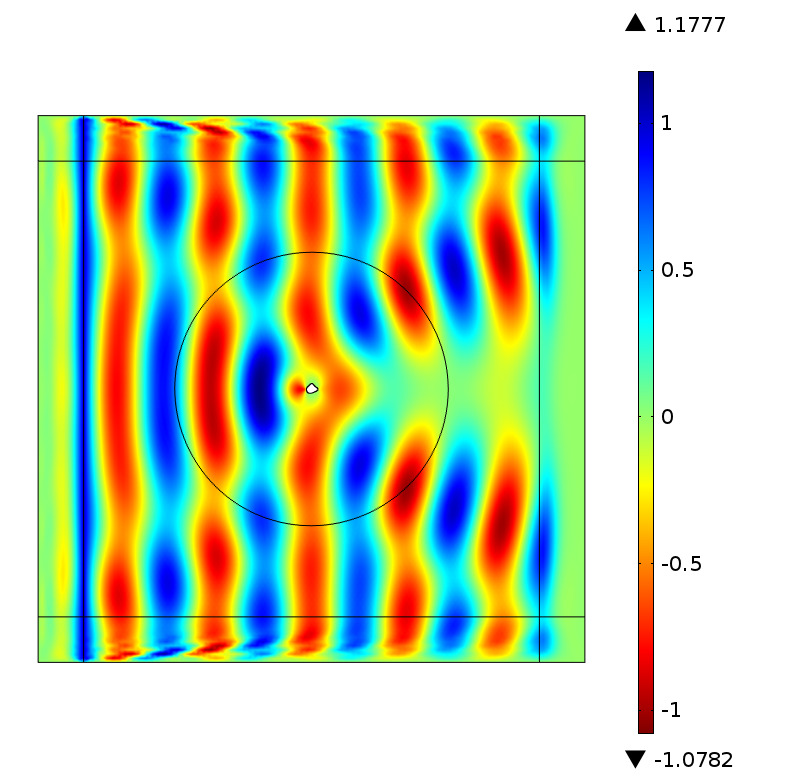}
\end{center}
\caption{Plane wave generated by displacing the boundary of the PML region on the left by 1 unit.} \label{COMSOL_plane_wave_nohole_and_shadow}
\end{figure}

The evaluation of $Q_j$ relies on the solution of the model problems discussed above. For the sake of convenience, we assume that $\theta_0=0$. 
For sufficiently large fixed $\beta r$, the series (\ref{plane_wave_scattered_field}) and (\ref{arbitrary_source_field}) are Fourier series (on the circle shown in figure \ref{COMSOL_plane_wave_nohole_and_shadow} (right)), and their coefficients are  numerically evaluated in  the commercial package COMSOL. This is followed by solving the system (\ref{fourier_coeff_eqn}) for the intensities ${\cal Q}_j$ of the control sources. Note here, that the number of sources determines the value of $K$ in equation (\ref{fourier_coeff_eqn}).

The unperturbed plane wave propagating horizontally and the plane wave interacting with a  clamped scatterer are shown in figure \ref{COMSOL_plane_wave_nohole_and_shadow}, left and right, respectively. We note that perfectly matched layers (PML) are used on the exterior boundary of the computational domain to provide non-reflective boundary conditions. A shadow region is clearly visible behind the scatterer in figure \ref{COMSOL_plane_wave_nohole_and_shadow} (right).

To set up an active cloak,  seven control sources are positioned around the scatterer, as displayed in figure \ref{COMSOL_result} (left).   
This means we have seven Fourier coefficients, which is enough to sufficiently approximate series (\ref{plane_wave_scattered_field}) and (\ref{arbitrary_source_field}).

The resulting computation in figure \ref{COMSOL_result} (right) indicates that the unperturbed plane wave has emerged behind the scatterer, so the cloak successfully eliminates the shadow region as required.

 \vspace{-0.5cm}
 
\begin{center}
\begin{minipage}{6.45in}
\centering
 ~~~~~~~\raisebox{-0.5\height}{\includegraphics[width=4.1cm]{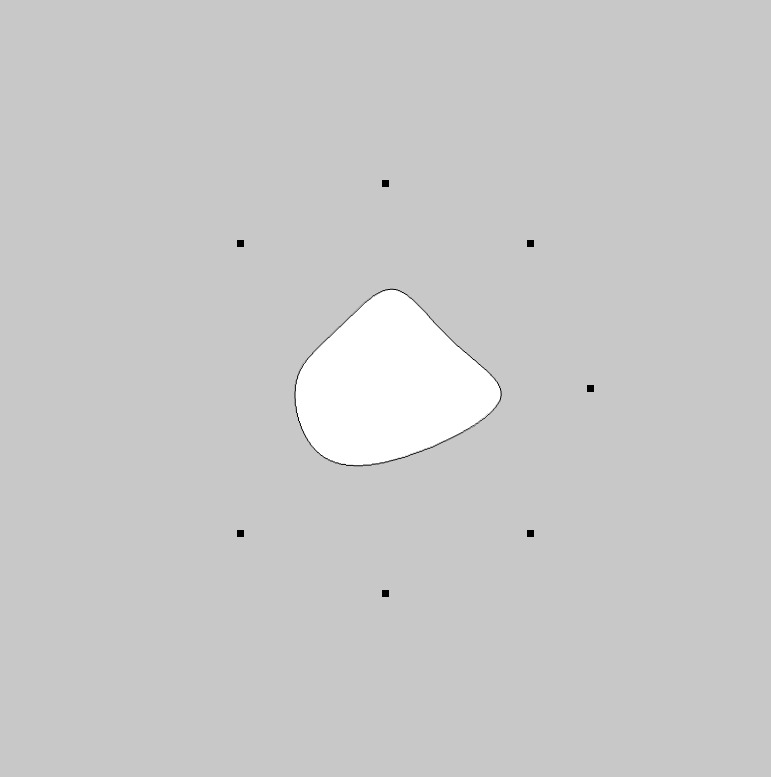}}
  ~~\raisebox{-0.5\height}{\includegraphics[width=7cm]{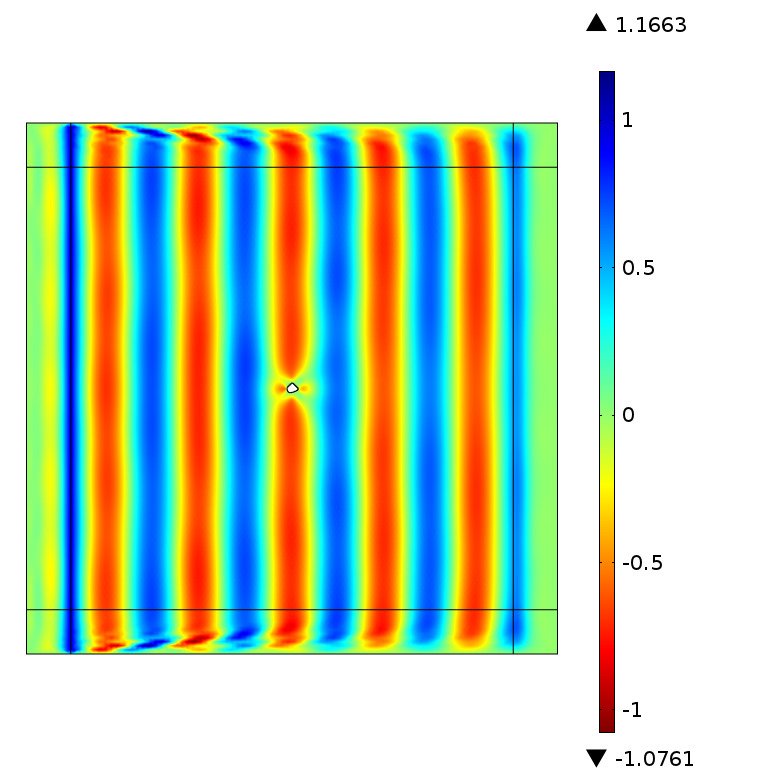}}
  \captionof{figure}
  {Left: Enlarged view of the scatterer and the locations of the control sources. Right: Effectively perfect cloaking achieved by the use of seven control sources. }
  \label{COMSOL_result}
\end{minipage}
    \end{center}

\begin{figure}[H]
\begin{center}
\includegraphics[width=6.5cm]{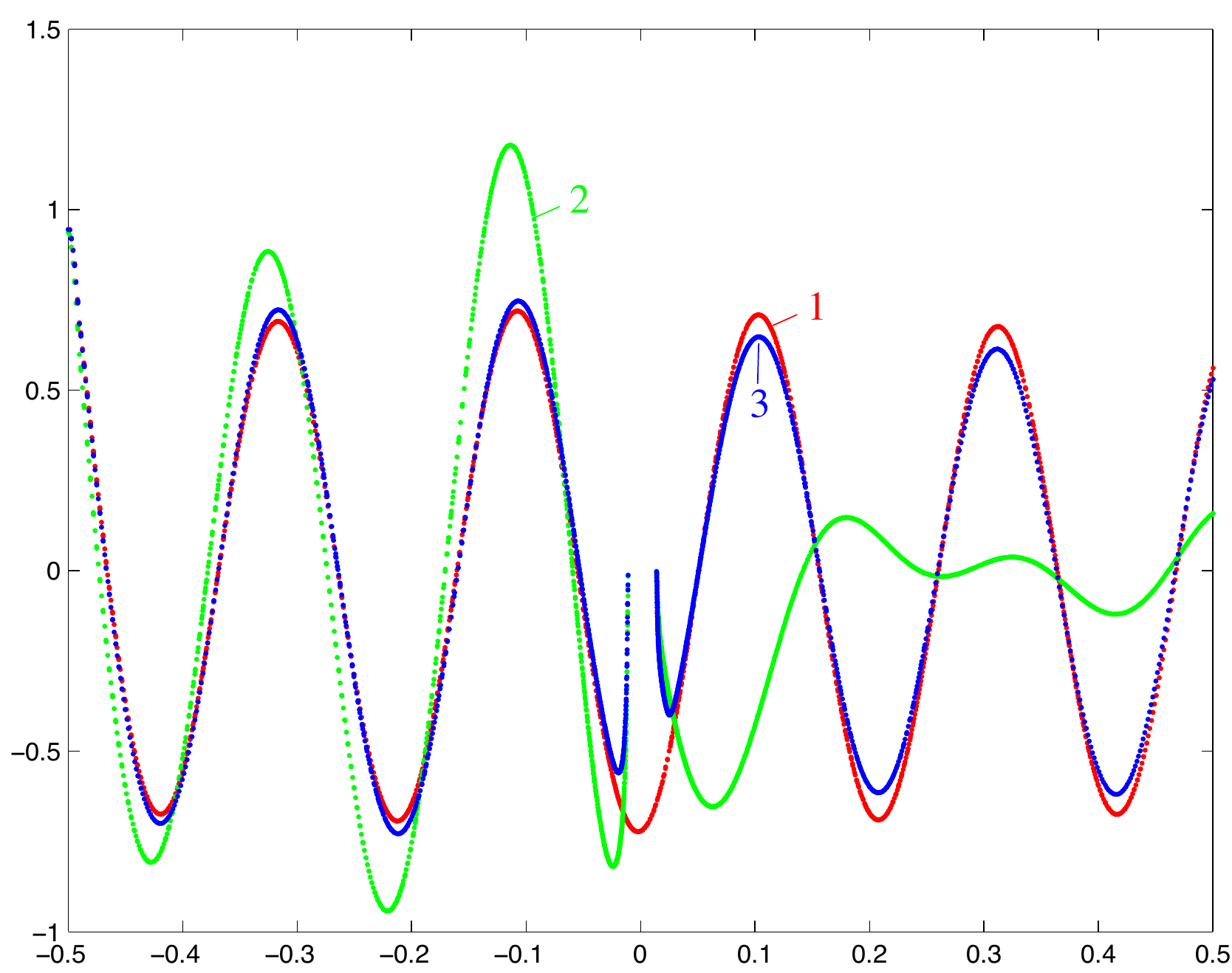}~~~
\includegraphics[width=6.5cm]{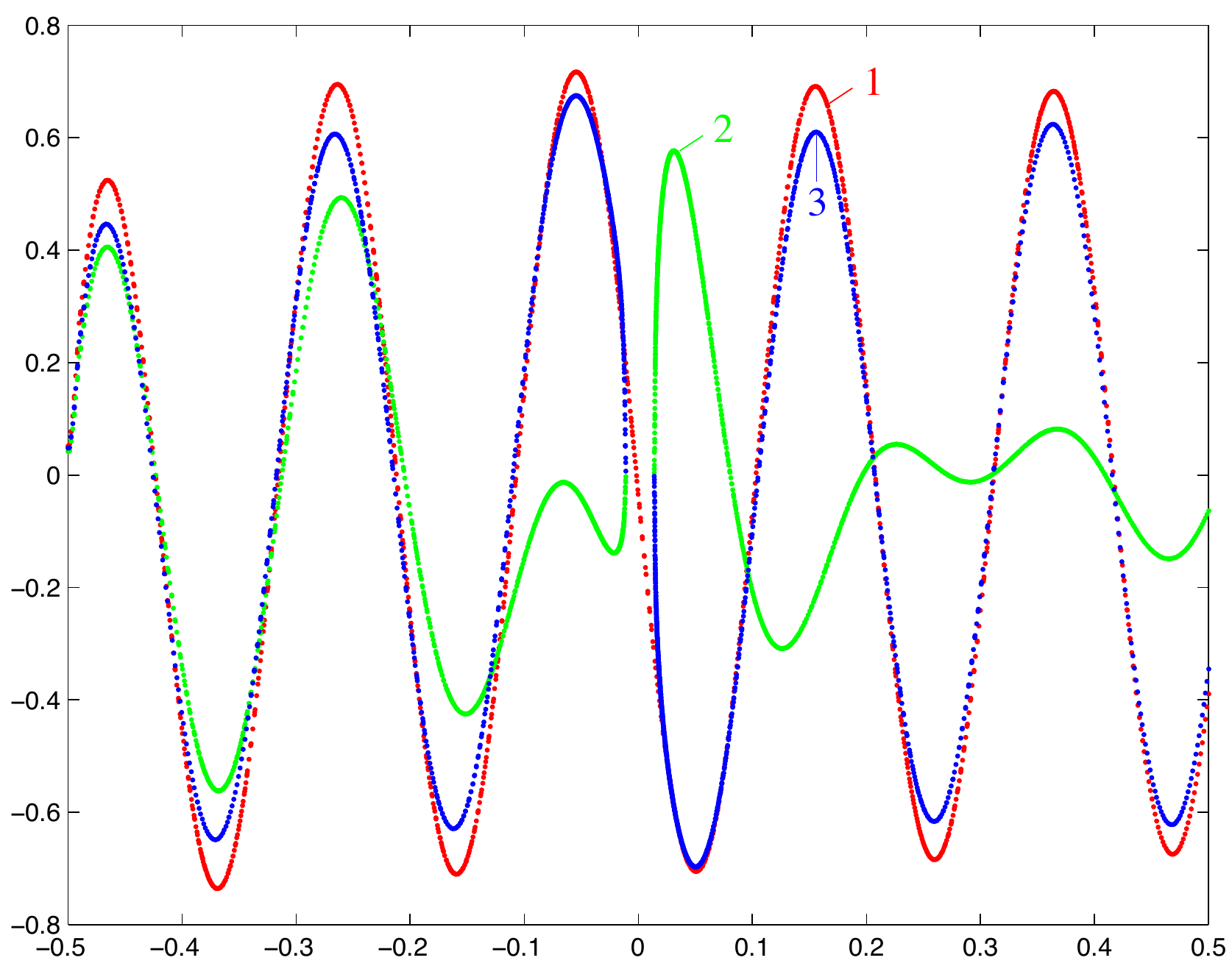}
\end{center}
\caption{The resulting total displacement field along the $x$-axis (left real, right imaginary) for plane wave propagation in a plate with no scatterer (figure \ref{COMSOL_plane_wave_nohole_and_shadow} left) in red (numbered as 1), with an arbitrarily shaped scatterer (figure \ref{COMSOL_plane_wave_nohole_and_shadow} right) in green (numbered as 2) and with the added active control sources (figure \ref{COMSOL_result} right) in blue (numbered as 3).} 
\label{along_the_x_axis}
\end{figure}

The efficiency of cloaking can be also shown by plotting the total displacement field along the $x$-axis of the plate (see figure \ref{along_the_x_axis}). Data set 2 (green) shows the real (left) and imaginary (right) parts of the field in a plate where there is an arbitrarily shaped scatterer with no control sources surrounding it. The presence of the shadow region is prominent in figure \ref{along_the_x_axis} (left) as the real part of the displacement field decreases dramatically behind the location of the scatterer. We also see no sinusoidal features of the imaginary part of the field from figure \ref{along_the_x_axis} (right).  Data set 1 (in red), shows the real (left) and imaginary (right) parts of the field in a plate with no scatterer, whilst data set 3 (in blue), shows that for an arbitrarily shaped inclusion with the seven surrounding point sources discussed earlier. 
It is clear that data set 3 lies within close proximity to data set 1, indicating that the incident plane wave has been reconstructed in the plate, and no scattering is apparent from the inclusion. Note that the small discontinuities in data sets 2 and 3 near the origin along the $x$-axis are due to the location of the inclusion and the boundary conditions imposed upon it.

\section{Concluding Remarks}

In this paper we have demonstrated an efficient algorithm of active cloaking for flexural waves scattered by rigid inclusions. The approach has been developed for the fourth-order differential operator, where the Green's function is represented as a normalised difference of the Helmholtz and modified Helmholtz Green's functions. The multipole expansion of the scattered field has been analysed in detail and solutions of model problems represent Green's function for a plate with a rigid inclusion and the plane wave scattered by  this rigid inclusion. The shape of the inclusion determines the multipole coefficients.

Furthermore, an efficient asymptotic algorithm allows for leading multipole terms in the scattered field to be cancelled by appropriate tuning of several point sources placed around the inclusion. The result is the suppression of the shadow region behind the inclusion, as required for the invisibility cloak.

The algorithm presented in the paper is generic and extends to other types of boundary conditions as well as arrays of defects rather than a single scatterer. 

If inclusions are arranged in a periodic array, as a diffraction grating, then the distribution of sources seen in figure \ref{COMSOL_result} (left) 
may contribute to a ``transparent'' grating,
where, after recalculating the required intensities of the sources, a plane wave is successfully transmitted through the grating. 

\section*{Acknowledgment}

J. O'Neill would like to greatly acknowledge the support from the EPSRC through the grant EP/L50518/1. R.C. McPhedran and N.V. Movchan acknowledge the financial support of the European Commission's Seventh Framework Programme under the contract number PIAPP-GA-284544-PARM-2.  R.C. McPhedran also acknowledges support from the Australian Research Council through its Discovery Grants Scheme. We would like to also thank Dr. D.J. Colquitt for his invaluable help with COMSOL.

\section*{Appendix}
We wish to determine the small frequency expansions ($\beta\rightarrow 0$) of the scattering matrix occurring in (\ref{gen_solns_in_terms_of_Wronskians}), which we will denote by ${\cal S}_n$. Using Mathematica, the expansions for the elements of the monopole matrix  are
\begin{eqnarray}
{\cal S}_0[1,1](\beta r_c)&=&-1+\frac{i}{\pi}\left[1+2\log(2)^2 
+2\gamma (-1+\gamma-\log(4))+\log(4) 
+2(-1+2\gamma+\log(\beta r_c/4)) \right.\nonumber \\
& &\left.\log(\beta r_c)\right](\beta r_c)^2+O(\beta r_c)^4,
\label{s011_expansion}
\end{eqnarray}
\begin{eqnarray}
{\cal S}_0[1,2](\beta r_c)&=&1-\frac{i}{2\pi}\left[2+4\gamma^2+\gamma  (-4-2 i\pi -8\log(2))+i\pi(1+\log(4))+(1+\log(2))\log(16)\right.\nonumber\\
 & &\left.+2\log(\beta r_c/4)) (-2+4\gamma-i\pi-4\log(2)+2\log(\beta r_c))\right](\beta r_c)^2+O(\beta r_c)^4,
\label{s012_expansion}
\end{eqnarray}
\begin{eqnarray}
{\cal S}_0[2,1](\beta r_c)&=&-\frac{2 i}{\pi}+\frac{1}{\pi^2}\left[-2-3 i\pi -4\log(2)^2+\gamma  (4-4\gamma+2 i\pi +8\log(2))-(4+2i\pi)\log(2)\right.\nonumber\\
 & &\left.+\log(\beta r_c) (4-8\gamma+2i\pi+8\log(2)-4\log(\beta r_c))\right](\beta r_c)^2+O(\beta r_c)^4,
\label{s021_expansion}
\end{eqnarray}
and
\begin{eqnarray}
{\cal S}_0[2,2](\beta r_c)&=&-\frac{2 i}{\pi}+\frac{1}{\pi^2}\left[-2-4\gamma^2+ \pi(-4 i+\pi) -4\log(2)^2+\gamma  (4+4 i\pi +8\log(2))\right.\nonumber\\
 & &\left.-(4+4i\pi)\log(2)+\log(\beta r_c) (4-8\gamma+4i\pi+8\log(2)-4\log(\beta r_c))\right]\nonumber\\
 & &(\beta r_c)^2+O(\beta r_c)^4.
\label{s022_expansion}
\end{eqnarray}
The correction terms in these expressions to the leading order term go to zero as $[ \beta r_c \log(\beta r_c)]^2$.

There are two sets of dipole terms, corresponding to $n=1$ and $n=-1$. For $n=1$, their first two terms of each are 
\begin{equation}
{\cal S}_1[1,1](\beta r_c)=-\frac{\pi}{4 i\gamma+\pi+4 i\log(\beta r_c/2)}-\frac{2 i\pi (\gamma+\log(\beta r_c/2))(\beta r_c)^2}{(-4\gamma+i\pi-4 \log(\beta r_c/2))^2},
\label{first_2_terms_s111}
\end{equation}
\begin{equation}
{\cal S}_1[1,2](\beta r_c)=\frac{\pi}{4 i\gamma+\pi+4 i\log(\beta r_c/2)}+\frac{(\pi \beta r_c)^2}{2(\pi+4i \gamma+4i \log(\beta r_c/2))^2},
\label{first_2_terms_s112}
\end{equation}
\begin{equation}
{\cal S}_1[2,1](\beta r_c)=\frac{-2}{4 \gamma-i\pi+4 \log(\beta r_c/2)}+\frac{i\pi( \beta r_c)^2}{(i\pi-4\gamma-4 \log(\beta r_c/2))^2},
\label{first_2_terms_s121}
\end{equation}
and
\begin{eqnarray}
{\cal S}_1[2,2](\beta r_c)&=&-\frac{2 i}{\pi}+\frac{1}{\pi^2}\left[-2(1+2\gamma^2+\gamma(-2-i\pi-4\log(2))\right. \nonumber\\
& &\left. +i\pi(1+\log(2))+\log(4)+\log(2)\log(4))\right. \nonumber \\
 & & \left.+\log(\beta r_c) (4-8\gamma+2i\pi-4\log(\beta r_c/4)) \right](\beta r_c)^2.
\label{first_2_terms_s122}
\end{eqnarray}

\end{document}